\documentclass[11pt]{article}
\pdfoutput=1
\usepackage{jheppub} 
\usepackage{enumerate}
\usepackage{amsmath,amssymb,epsfig,cancel}
\usepackage{color}
\usepackage{subfig}
\usepackage{xcolor}
\usepackage{graphicx}
\usepackage{caption}
\usepackage{braket}
\usepackage{hyperref}
\usepackage{float}
\usepackage[english]{babel}
\usepackage{tikz}
\usetikzlibrary{arrows}
\usetikzlibrary{matrix}

\title{Infinite Linear Quivers and Continuous Rank Functions} 
\author{Jeroen van Gorsel}

\affiliation{Department of Physics, Swansea University, Swansea SA2 8PP, United Kingdom}

\emailAdd{
 jeroen.van.gorsel@gmail.com} 

\abstract{We introduce continuous rank functions for field theories with linear quivers, and demonstrate they describe infinitely long quivers. 
In the context of AdS/CFT, these `continuous quivers' have well-defined supergravity dual descriptions with an infinite number of smeared NS5 and D-branes. These continuous rank functions generalise the supergravity solutions one can consider to be the dual descriptions of quiver field theories. We demonstrate this construction explicitly in the context of the half-BPS classes of AdS$_5$ and AdS$_7$ geometries, where we construct examples of a continuous parabolic and sinusoidal rank function.\\
\\
\\[10pt]
 } 

\subheader{}

\begin{document}
\def\Tr{{\textrm{Tr}}}


\maketitle 

\newpage
\section{Introduction}
Since the increadible advances in the understanding of maximally supersymmetric conformal field theories (SCFTs) like $\mathcal{N} = 4$ SYM in the context of AdS/CFT\cite{Maldacena:1997re}, a lot of progress has been made in the study of half-supersymmetric SCFTs. Such half-BPS field theories can be constructed as the low energy fluctuations of Hanany-Witten set-ups \cite{Hanany:1996ie} containing NS5-branes with stacks of various numbers of D$_p$-branes suspended between them 
\cite{Hanany:1997gh, 
Aharony:1997bh, Aharony:1997ju, 
Gaiotto:2009we, 
Gaiotto:2008ak}, 
giving a $p$-dimensional field theory as the fluctuations between the NS5-branes freeze out at low energies. Additional stacks of D$_{p+2}$-branes can be added, giving rise to flavour groups. Strings going between these different stacks of colour and flavour branes give rise to hypermultiplets that transform under the bifundamental representation of the $SU(N_F)$ and $SU(N_C)$ flavour and colour groups, leading to a quiver gauge theory. 

At the conformal fixed point, these theories are thought to have a holographic dual description in terms of half-supersymmetric AdS$_{p+1}$ supergravity solutions \cite{
Cremonesi:2015bld, Gaiotto:2014lca, Bobev:2016phc, 
DHoker:2016ujz, DHoker:2017mds, DHoker:2017zwj, 
Gaiotto:2009gz, 
DHoker:2007zhm, DHoker:2007hhe}.
Various recent works have shown particularly interesting properties of supergravity solutions corresponding to both infinite \cite{
Lozano:2018pcp, 
Lozano:2016kum, Nunez:2018qcj, 
Lozano:2016wrs, 
Filippas:2019ihy}, 
and continuous \cite{Filippas:2019puw} rank functions in these SCFTs.
In this paper show how to make sense of such continuous rank functions, as describing well-defined infinitely long quivers. We will focus primarily on the half-BPS AdS$_5$ and AdS$_7$ cases. For these set-ups the ranks of the different colour and flavour groups have to be related as
\begin{equation}
F_n = 2N_n - N_{n+1} - N_{n-1},
\end{equation}
so that the flavour groups can be thought of as the second derivative of the colour groups. In Appendix \ref{App_AdSGeometries} we briefly comment on the AdS$_4$ and AdS$_6$ cases. 
This paper is organised as follows:

\begin{itemize}
\item In section \ref{sec_BraneSetUps} we introduce the half-BPS Hanany-Witten brane set-ups that give rise to these quiver field theories. 

\item In section \ref{sec_InfiniteQuiversAndContinuousRanks} we discuss the supergravity limit for long quivers with large ranks for the gauge and flavour groups. We review a conventional way of `scaling' a finite quiver to one where the supergravity description is a valid approximation. We then show how one can instead define an infinite quiver starting from the flavour groups and using the above consistency condition to fix the ranks of the gauge groups. 

Using this definition for infinite quivers we define continuous rank function. We calculate the Page and central charges for quivers with continuous rank functions, and show the results agree with infinitely long discrete rank functions. We discuss two examples of continuous rank functions in detail: a parabolic one with $R(z) = [n + \frac12 f (1-z)]z$ and a sinusoidal one where $R(z) = A \sin \omega z$.

\item Another way of thinking about these continuous rank functions is that the flavour branes on the geometries are `smeared'. In section \ref{sec_SmearedBranes} we approach the idea of continuous rank functions from this angle, and show this leads to the same conclusions as in the previous section.

\item In section \ref{sec_AdS5andAdS7} we discuss in more detail how these ideas apply directly to the half-BPS classes of AdS$_5$ and AdS$_7$ geometries. We show how to explicitly construct solutions corresponding to infinite quivers with continuous rank functions. The way this can be generalised to the half-BPS AdS$_6$ and AdS$_5$ classes is not directly obvious, we commend on this in Appendix \ref{App_AdSGeometries}. 

\item In section \ref{sec_Conclusion} we briefly discuss the field theory interpretation of these continuous rank functions. 

\end{itemize}

\section{Brane Set-ups for Quiver SCFTs}\label{sec_BraneSetUps}
\begin{figure}[t!]
\centering
\subfloat[\small \normalsize]{\label{fig_braneSetup}
     \includegraphics[width=0.6\textwidth]{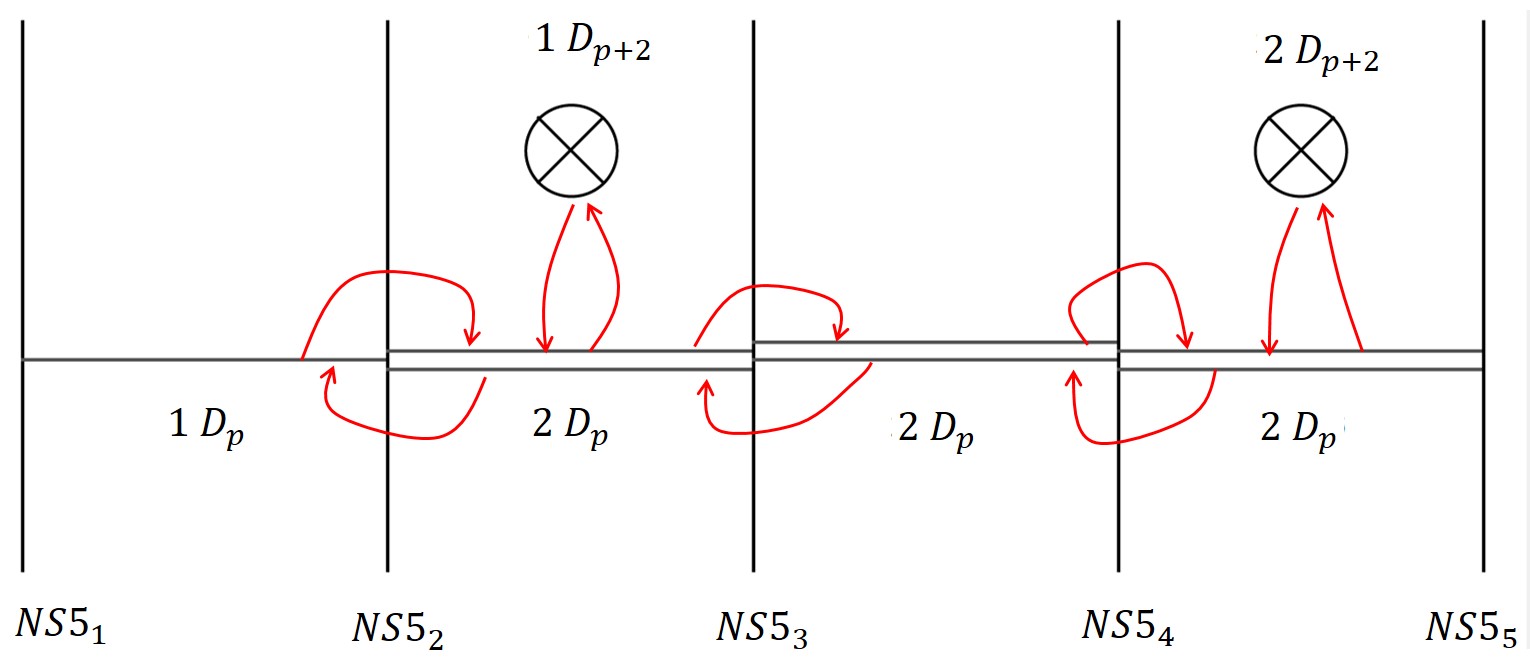}}
\\    
\centering
\subfloat[\small \normalsize]{\label{fig_rankfunction}
     \includegraphics[width=0.3\textwidth]{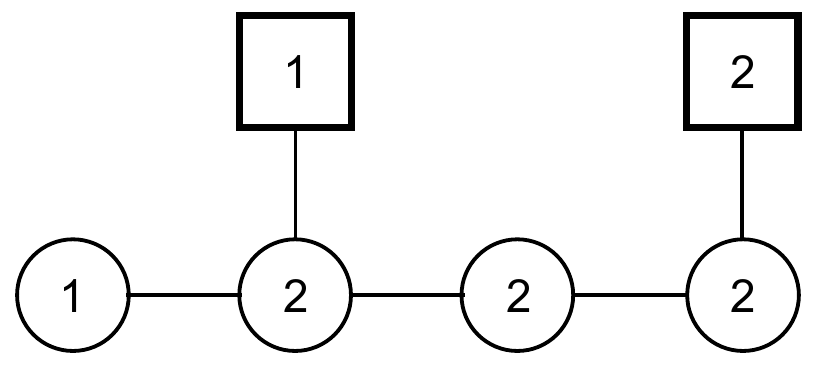}}
\subfloat[\small \normalsize]{\label{fig_functions}
    \includegraphics[width=0.3\textwidth]{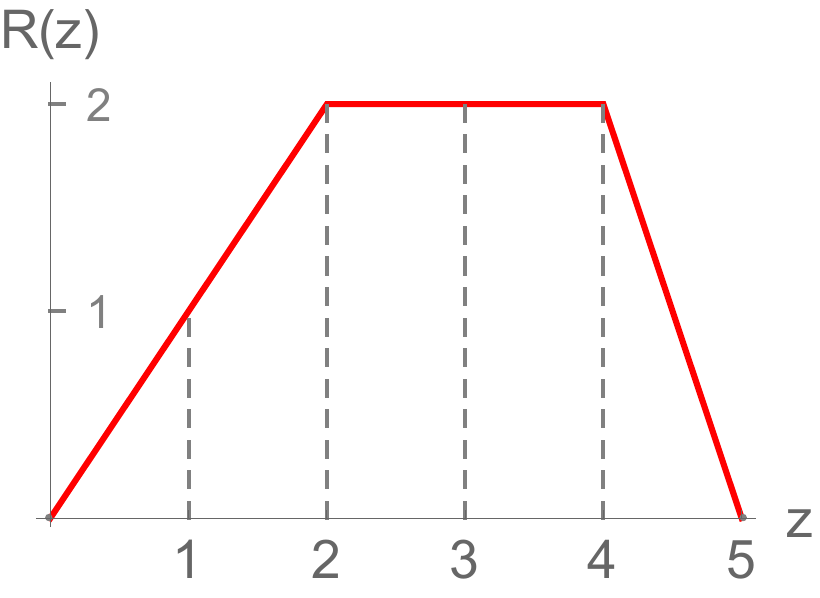}}    
\caption{Example of a Hanany-Witten D6-D8-NS5 brane set-up. \textbf{(a)} Strings going between the branes, indicated by red arrows, give rise to hypermultiplets. \textbf{(b)} Quiver diagram corresponding to the brane setup, circles indicate colour groups and boxes indicate flavour groups. \textbf{(c)} The corresponding rank function for this brane set-up.} \label{fig_TotalQuiverExample}
\end{figure}

Hanany-Witten set-ups \cite{Hanany:1997gh} with stacks of D$_p$-branes suspended between NS5-branes (see figure \ref{fig_TotalQuiverExample}) give rise to half-supersymmetric $p$-dimensional SCFTs. The literature on this topic is vast, and it is difficult to do it justice in this short introduction. We present these constructions for various values of $p$ in more detail in section \ref{sec_AdS5andAdS7} and Appendix \ref{App_AdSGeometries}, with references \cite{Cremonesi:2015bld}-\cite{Lozano:2019emq}. These theories have a gauge group
\begin{equation}\label{eq_QuiverProductOfGaugeGroups}
G = SU(N_1) \times SU(N_2) \times \ldots \times SU(N_{n-1}) \times SU(N_n).
\end{equation}
Each of the $n$ $D_p$-branes gives rise to a vector multiplet, transforming under the adjoint representation of the corresponding $SU(N_i)$ gauge group, describing the fluctuations of the $D_p$-brane stacks. In addition, there will be $n$ hyper multiplets, that transform as singlets under the $SU(N_i)$, and encode the fluctuations in the distance between the different NS5-branes.

The \emph{quiver} structure in these set-ups comes from strings going between the different D-branes, and give rise to $n-1$ bifundamental hypermultiplets $\Psi$ that transform under the fundamental representation of one of the $SU(N_i)$ gauge groups, while their adjoint $\bar{\Psi}$ transforms under the anti-fundamental representation of a consecutive gauge group $SU(N_{i\pm1})$. 

Flavours can be introduced by adding additional $D_{p+2}$-branes that do not extend between the NS5-branes, but extend on three additional directions perpendicular to the NS5-branes. As these $D_{p+2}$-branes are larger and heavier, their dynamics will freeze out. At low energies strings extending from one of the $D_p$-branes to these heavier $D_{p+2}$-branes will introduce additional flavour indices on the hypermultiplets. To keep track of all these different hypermultiplets one can summarise the colour and flavour groups in a quiver diagram, as is shown for an example in figure \ref{fig_rankfunction}.

\subsection{Consistency Condition at the Conformal Fixed Point}
In the remainder of this article we will focus in particular on the cases where $p= 4,6$\footnote{For $p = 3,5$ the consistency condition is different, and some of the results mentioned here and used throughout this paper are no longer valid. We comment on this in more detail in Appendix \ref{App_AdSGeometries}.}! 

At the conformal fixed point the worldvolume theories are expected to be holographically dual to an AdS$_{p+1}$ supergravity solution \cite{Maldacena:1997re}. 
For these 4d and 6d SCFTs the ranks of the different gauge and flavour groups ($N_n$ and $F_n$ respectively) cannot be arbitrary at the conformal fixed point\footnote{For $p=4$ this condition will be equivalent to the vanishing of the beta-function. For $p=6$ this condition ensures the vanishing of gauge anomalies in the 6d SCFT.}, but have to be related to each other such that 
\begin{equation}\label{eq_CFConsistencyCondition}
F_n = 2N_n - N_{n+1} - N_{n-1}.
\end{equation}
This means the flavour groups act as the second derivative of the colour groups. As a result, we the colour groups and flavour groups of the quiver are fixed in terms of one another at the conformal fixed point. This property is central for the way we define infinite quivers in section \ref{subsec_StartingFromFlavours}.

It is expected that these AdS$_{p+1}$ backgrounds can arise as the near horizon limit of the brane set-up when the distances between the different NS5-branes is go to zero. 
To satisfy the above consistency condition for a quiver with particular ranks for the colour groups, 
the AdS geometry carries the information of the NS5 and D$_p$-branes in it the fluxes on the background. One can include the $D_{p+2}$ flavour branes in the AdS geometry by letting them backreact on the near-horizon `colour' geometry. The backreaction of these flavour branes will deform the supergravity geometry, which will now satisfy the equations of motion following from the action
\begin{equation}\label{eq_TypeIIBIWZAction}
S_{Type\;II} +  \delta^{9-(p+2)}(\vec{x} - \vec{x}_{D_{p+2}}) S_{BIWZ}
\end{equation}
where $S_{BIWZ}$ is the Born-Infeld action for the $D_{p+2}$-flavour branes, amended with the appropriate Wess-Zumino term. Since these flavour branes are localised this action has to be multiplied with a delta function. The BIWZ action is of the form
\begin{equation}
S_{BIWZ} = -T_p \int d^{p+3}\sigma e^{-\Phi} \sqrt{-\text{det} ( g_{ab} - \mathcal{F}_{ab} )}
+ T_p \int d^{p+1}\sigma\;\mathcal{C}\wedge e^{-\mathcal{F}}
\end{equation}
with $\mathcal{F}_{ab}$, $\Phi$, $g_{ab}$ and $B_{ab}$ the pullbacks of the supergravity fields induced by the background of the colour branes. The coupling to the RR-fields of the colour background is given by the WZ-part of the action. 

The additional BIWZ term in the action (\ref{eq_TypeIIBIWZAction}) modifies the equations of motion. In particular, the Bianchi identities for the RR-fluxes that couple to the flavour brane will now be of the form
\begin{equation}
dF_{8-(p+2)} = N_f\;\delta^{9-(p+2)}(\vec{x} - \vec{x}_{D_{p+2}}).
\end{equation}
Since the $D_{p+2}$-flavour branes will act as sources for the $F_{8-(p+2)}$-flux. Everywhere in the geometry the Bianchi identities $dF_{8-(p+2)} = 0$ will be satisfied, except at the points where these flavour-branes are located.\\
\\
If we demand that the consistency condition (\ref{eq_CFConsistencyCondition}) is satisfied, we only need the ranks of the gauge groups to define a particular quiver (for $p=4,6$), the flavour groups then follow from the consistency condition. Since the colour ranks contain all the information of the brane set-up, we would expect the dual AdS geometry, including the backreaction of the flavour branes, to be completely determined by this rank function.
\begin{equation}
R_i = \left( N_1, N_2, \ldots, N_{n-1}, N_n \right)
\end{equation}
This is indeed the case for the AdS$_7$ backgrounds that we will introduce in more detail in section \ref{subsec_AdS7}, and the AdS$_5$ backgrounds in section \ref{subsec_AdS5}. Given a particular rank function, that should now be thought of as a piece-wise continuous linear function $R(z) \in C^{(0)}$ (see figure \ref{fig_functions}) the dual geometries are completely determined.
 
In general all coefficients in the metric, NS- and RR-fluxes are given in terms of the derivatives of one or multiple functions $\mathcal{F}_i(\Sigma_2)$, that has to satisfy a particular differential equation for the geometry to satisfy the BPS equations. The rank function is typically related to the boundary condition for this differential equation in all these half-BPS backgrounds. We discuss the classes of half-BPS AdS$_7$ and AdS$_5$ backgrouds in more detail in section \ref{sec_AdS5andAdS7}. See Appendix \ref{App_AdSGeometries} for the AdS$_4$ and AdS$_6$ cases.

\section{Infinite Quives and Continuous Rank Functions}\label{sec_InfiniteQuiversAndContinuousRanks}
One way of obtaining an SCFTs for which the supergravity description is valid, is by considering very long quivers with large numbers for the colour and flavour groups. 
Here we will first describe the conventional approach of obtaining long quivers, by `scaling' an initial shorter quiver to one for which the supergravity description is valid.
 
We then introduce a different way of defining infinitely long quivers. We show how by first defining an arbitrary number of flavour groups $F_n$, and then using the consistency condition to fix the ranks of all the colour groups, one can define a quiver of arbitrary length. In the limit the the ranks of the colours groups are everywhere much larger than the flavour groups $N_n/F_n \to \infty$ this gives rise to continuous rank functions. We will illustrate this in detail for two particular cases, a parabolic rank function (section \ref{subsec_ParabolicQuiver}), and a sinusoidal rank function (section \ref{subsec_SinusoidalQuiver}).

\subsection{Scaling a Finite Quiver}\label{subsec_ScalinginiteQuiver}
\begin{figure}[h!]
\centering
\subfloat[]{\label{fig_RankScalingRank}
     \includegraphics[width=0.3\textwidth]{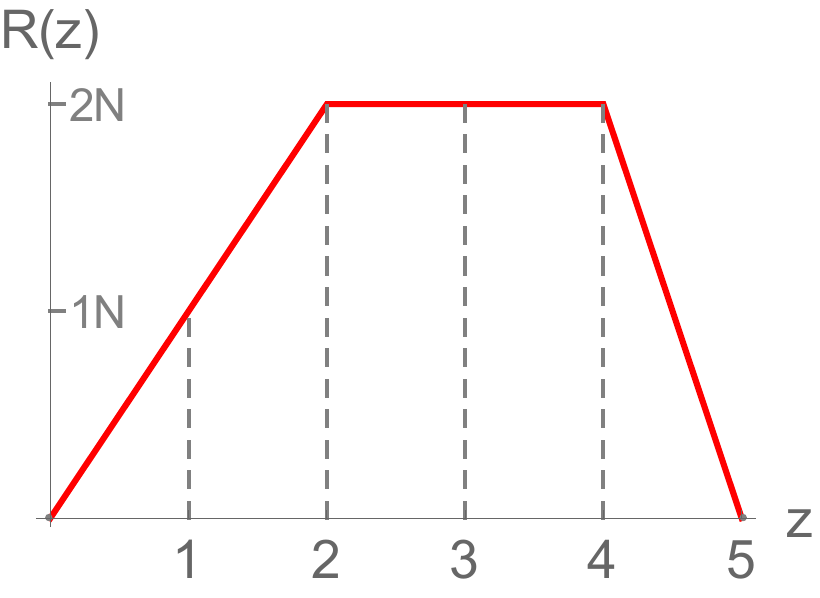}}
\subfloat[]{\label{fig_RankScalingQuiver}
    \includegraphics[width=0.17\textwidth]{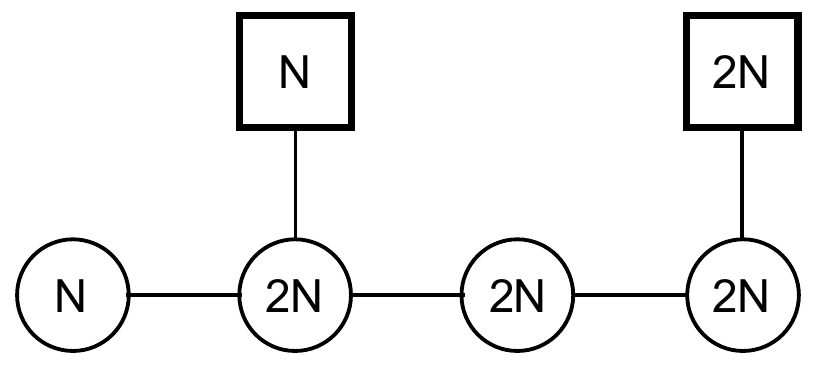}} 
\caption{The quiver and rank function of figure \ref{fig_TotalQuiverExample}, now rescaled with $R(z) \to NR(z)$.} \label{fig_RankScaling}
\end{figure}

For the supergravity description to be a valid approximation of the dual SCFT the string coupling everywhere needs to be weak so we do not have to lift the ten-dimensional background to M-theory, furthermore the curvature needs to be small so we do not have to include stringy corrections of higher curvature terms in the effective supergravity action. \textcolor{red}\\
\\
In both the AdS$_5$ backgrounds of eq. (\ref{eq:10dGaiottoMaldacenaNS}) and AdS$_7$ backgrounds of eq.(\ref{eq:TomasielloGeometryGeneral}), we can scale the rank functions $R(z) \to N R(z)$ and obtain a different supergravity solution. 
This scaling cancels out in the metric, but  multiplies the RR-fluxes with the same prefactor $N$, while the string coupling scales as $N^{-1}$. This scaling thus increases the number of $D_p$ colour and $D_{p+2}$-flavour branes, while simultaneously decreases the string coupling $g_s^2$, keeping the number of NS5-branes fixed. By starting with a finite quiver this way and taking $N \to \infty$ we can take $g_s^2 \to 0$, while keeping $g_s^2 N$, as well as the overall shape of the quiver diagram fixed, see figure \ref{fig_RankScaling}.
\begin{equation}\label{eq_SugraRescaling}
e^{2\phi} \to \frac{1}{N} e^{2 \phi}, \hspace{2cm}
F_p \to N F_p, \hspace{2cm}
F_{p+2} \to N F_{p+2}.
\end{equation}
This limit thus decreases the closed string loop corrections (as we send $g_s^2 \to 0$). 
Note that since the flavours and colours are related to one another by the consistency condition (\ref{eq_CFConsistencyCondition}) this particular scaling multiplies the ranks of both the colour and flavour groups by $N$, while keeping their fixed\footnote{This in combination with $\lambda = g_{YM}^2 N \to \infty$ corresponds to taking the Veneziano limit, where $F_n \to \infty$ and $F_n/N_n$ is kept fixed (where $N_n$ and $F_n$ indicate the rank of the $n$-th colour and flavour group respectively).}. 

\begin{figure}[h!]
\centering
\subfloat[]{\label{fig_LengthScalingQuiver}
     \includegraphics[width=0.3\textwidth]{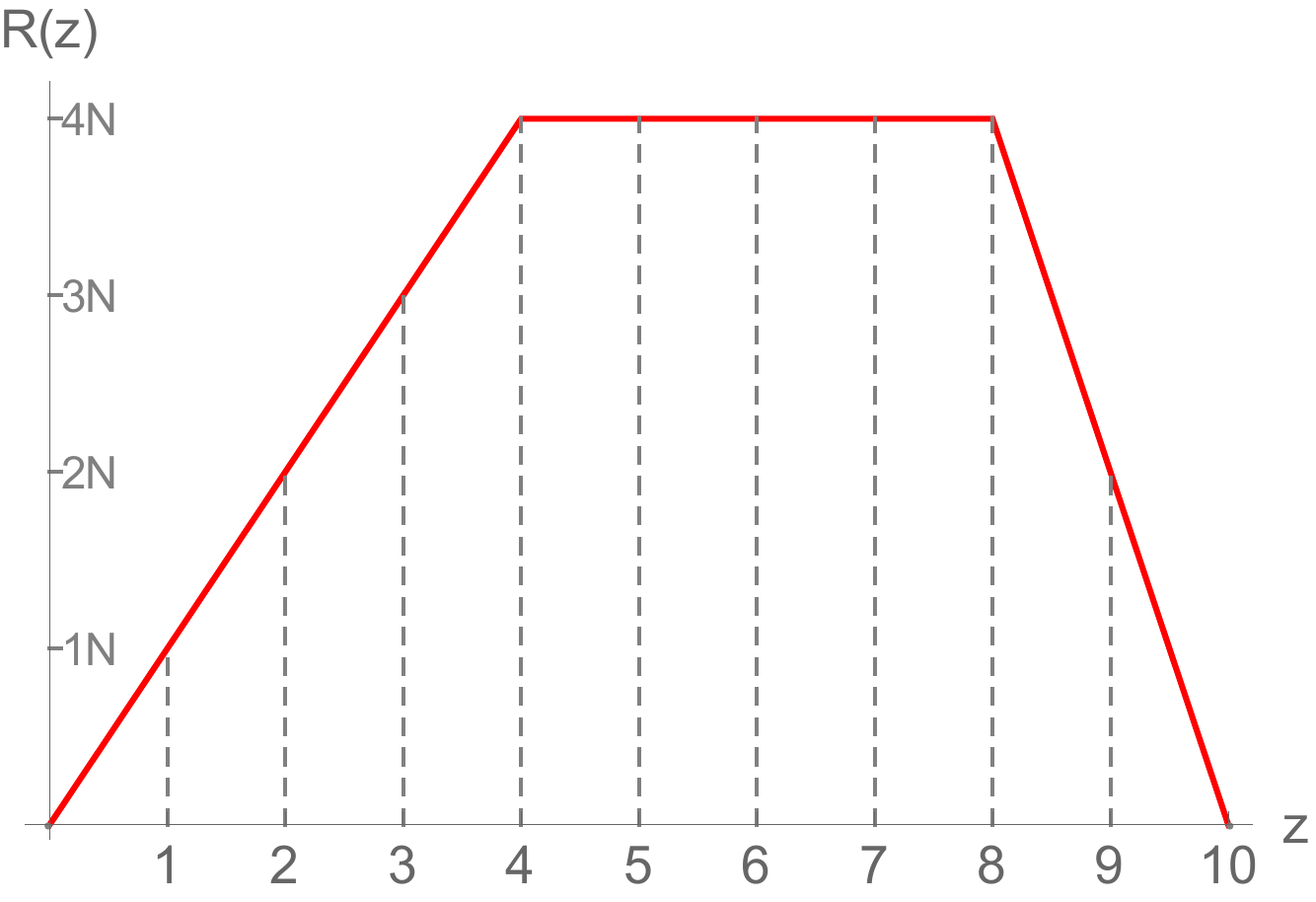}}
\subfloat[]{\label{fig_LengthScalingRank}
    \includegraphics[width=0.4\textwidth]{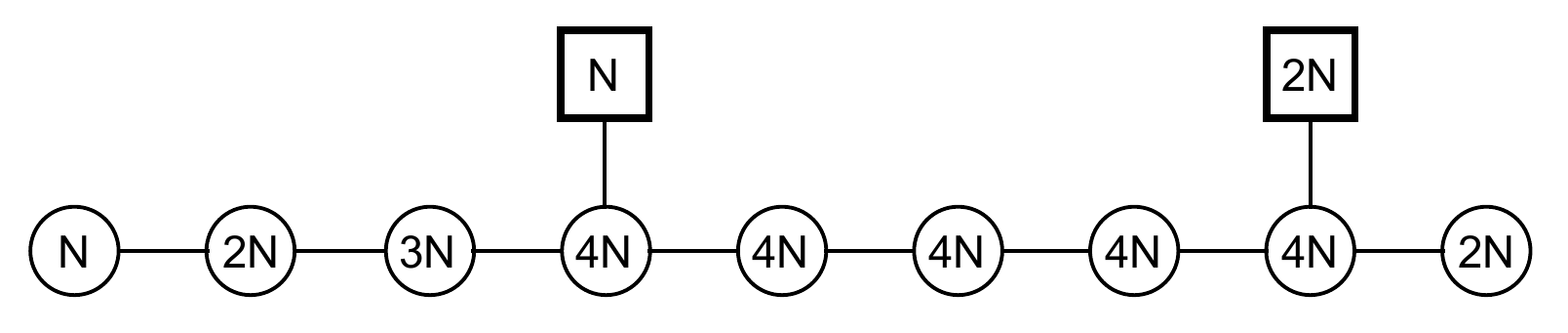}} 
\caption{The quiver and rank function of figure \ref{fig_TotalQuiverExample}, now with the combined scaling $R(z) \to NR(z)$ and $z \to \omega z$. Note the flavour groups stay the same as those in figure \ref{fig_RankScaling}.} \label{fig_LengthScaling}
\end{figure}

We can alternatively scale the $z$-direction in the supergravity solutions of eqs.(\ref{eq:10dGaiottoMaldacenaNS}) and (\ref{eq:TomasielloGeometryGeneral}). If we combine this with the $R(z) \to N R(z)$ scaling, we can increases the number of NS5- and D$_p$-colour branes in the corresponding set-up, while keeping the number of flavours-branes fixed, see figure \ref{fig_LengthScaling}. In addition to the string coupling getting suppressed, the internal direction on the space now grows, which reduces the curvature. 

By combining these two scalings one can start from a finite quiver - like the one in figure \ref{fig_TotalQuiverExample} - and repeatedly scale it to large $N$ and large number of NS5-branes, to obtain a very long quiver for which the supergravity description is valid. As a result of this scaling, the final quiver has only a few localised flavour groups $F_n$, for which we are in the Veneziano limit where $F_n \to \infty$ and $N_n \to \infty$, with$F_n / N_n$ kept fixed.

\subsection{Defining Quivers from their Flavour Groups}\label{subsec_StartingFromFlavours}
Let us now present a different way to define a large quiver. We do this by first defining all of the flavour groups $F_k \geq 0$, where $k$ is an arbitrary number giving the length of the quiver.

The 4d and 6d half-supersymmetric CFTs have the nice property that once all the flavour groups have been defined, the the ranks of the gauge groups $N_k$ are fixed by the consistency condition of eq. (\ref{eq_CFConsistencyCondition}),
\begin{equation}\label{eq_AnomalyCancelationCondition}
2 N_k - N_{k+1} - N_{k-1} = F_k.
\end{equation}
We let the rank function start with $N_{0} = 0$, and $N_1$ some arbitrary number for the first colour group in the quiver. Inserting this in the above condition then dictates the rank of the next colour group
\begin{equation}
N_1 = 2N_0 - F_0 - N_{-1} = 2N - F_0.
\end{equation} 
Having obtained the rank of the next colour group this way, one can repeat this process to recursively obtain the values for the consecutive rank of all the colour groups.
\begin{eqnarray}\label{eq_DefineQuiverFromAnomalyCancelation}
& N_2 =& 2N_1 - F_1 - N_{0} = 3N_1 - 2F_0 - F_1, \nonumber\\
& N_3 =& 2N_2 - F_2 - N_{1} = 4N_1 - 4F_0 - 2F_1 - F_2, \\
& & \ldots \nonumber
\end{eqnarray}
A quiver defined in this way will automatically terminate itself by giving $N_{\mathrm{last}} = 0$, for the last value of the rank function. This can be seen from the consistency condition, which for these theories implies the values of the flavour groups are proportional to the second derivative of the rank function
\begin{eqnarray}
& F_k &= (N_k - N_{k-1} ) - (N_{k+1} - N_k ),\\
&     &= \Delta N_k - \Delta N_{k+1} = \Delta^2 N_k.
\end{eqnarray}
As the values of the flavour groups have to be greater than or equal to zero everywhere, the rank function has to be concave, and will go back to zero eventually. For the quiver to become long want the second derivative of the rank function (and thus the ranks of the flavour groups $F_n$ have to be small compared to the colour groups $N_n$). 
This then implies that for every node of the quiver $F_n / N_n \to 0$, which means we are in the 't Hooft limit instead of the Veneziano limit. 

\subsection{Parabolic Quiver}\label{subsec_ParabolicQuiver}
\begin{figure}[b]
{
 \centering
 \subfloat[\small ]{
     \includegraphics[width=0.10\textwidth]{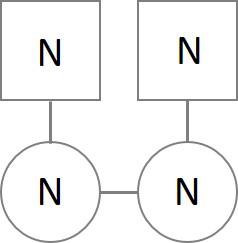}}\hspace{0.05\textwidth}
 \subfloat[\small ]{
    \includegraphics[width=0.3\textwidth]{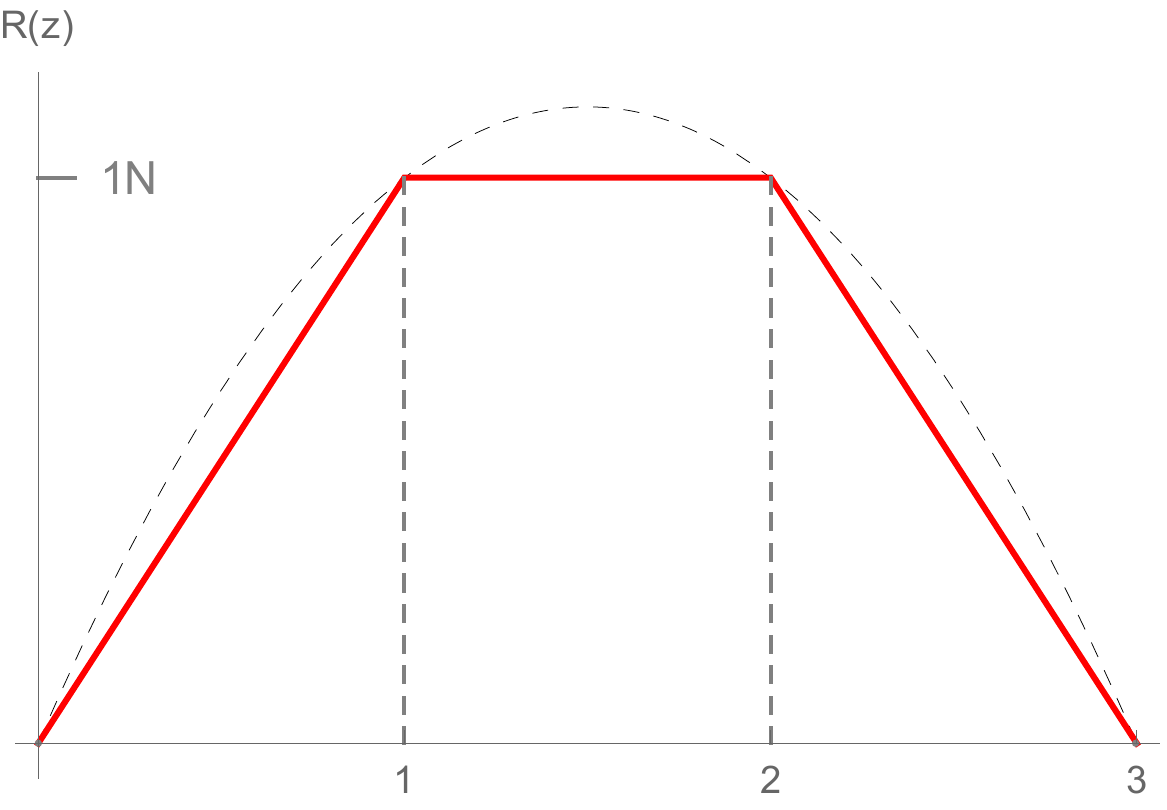}}   
\caption{(left) the quiver, and (right) the rank function (in red) for $n=1$ and $f=1$.}\label{fig_ParabolicQuiverTwoNodes}
}
\end{figure}

As an example we will start here with a quiver where all the flavour groups are equal to some arbitrary integer number
\begin{equation}
F_k = f.
\end{equation}
And choose the rank for the first gauge group on the quiver to be equal to some other arbitrary number $N_1 = n$. The ranks of the gauge groups will be
\begin{equation}\label{eq_ParabolicQuiverRank}
N_k = nk - \sum_{j=1}^{k} (j-1) f.
\end{equation}
If we set $N_1 = n$ we find from the consistency condition eq.(\ref{eq_AnomalyCancelationCondition}) that this will result in a rank function $R(z) = (2, 2)$, and the quiver will terminate after 2 colour nodes as can be seen in figure \ref{fig_ParabolicQuiverTwoNodes}.
For a generic quiver of this form, with $n$ and $f$ arbitrary numbers, the rank function will return to zero for $N_k$ when $k = 2n/f + 1$. We will first calculate the total number of branes and central charges associated with brane set-ups involving this quiver, using our results in Appendix \ref{sec_GeneralExpressions}. We then show that in the limit $n/f \to \infty$ the rank function asymptotes to that of continuous parabola (show with the dotted grey line in figure \ref{fig_ParabolicQuiverTwoNodes} ).\\
\\
The total number of $D_d$ and $D_{d+2}$ branes corresponding to this discrete quiver will be given by
\begin{eqnarray}
& N_{D_d} =& \sum_{k=1}^{2n/f}\left( kn - \sum_{j=1}^{k} (j-1)f \right) = n\left( \frac{2n^2}{3f^2} + \frac{n}{f} + \frac{1}{3} \right), \label{eq_NDd_DiscreteParabolicQuiver}\\
& N_{D_{d+2}} =& 2n. \label{eq_NDdp2_DiscreteParabolicQuiver}
\end{eqnarray}
Using this rank function in eq.(\ref{eq_ParabolicQuiverRank}) and the result from eq.(\ref{genericcc}) we can calculate the central charges of the corresponding $d=4$ and $d=6$ dimensional field theories as (with $P = 2n/f$)
\begin{equation}
c_{d=4} = \sum_{k=1}^{P} N_k^2  = n^2\left( \frac{2n^2}{3f^2} + \frac{n}{f} + \frac{1}{3} \right)^2 \label{eq_ccMG_ParabolicQuiver}
\end{equation}
For the $6d$ case we find 
\begin{eqnarray}\label{eq_ccCT_ParabolicQuiver}
& b_0 =& \frac{-1}{\frac{2n}{f}+1}\sum_{i=1}^{2n} \left( \frac{2n}{f}+1 - i\right) N_i, 
\hspace{1cm} = \frac{-n}{6f^2}(f+n)(f+2n),\\
& a_k =& k b_0 + \frac{N_k}{6} +\sum_{i=1}^k (k-i)N_i,
\hspace{1cm}=kb_0 + \frac{k^2}{24} \left( 4kn - f(k-1)^2 \right),\\
& b_k =& b_0 + \frac{N_k}{2} + \sum_{i=1}^{k-1} N_i,
\hspace{1cm}= b_0 + \frac{k}{12}\Large( 6kn - f(k-1)(2k-1) \Large),
\end{eqnarray}
giving us for the central charge
\begin{equation}\label{eq_6dccDiscreteParabolicQuiver}
c_{d=6} = -n^2\left( 
\frac{34 n^5}{315 f^5}+\frac{17 n^4}{45 f^4}+\frac{47 n^3}{90 f^3}+\frac{13 n^2}{36 f^2}+\frac{23 n}{180 f}+\frac{f}{2520 n}+\frac{7}{360} \right).
\end{equation}\\
\begin{figure}[t]
{
 \centering
 \subfloat[\small ]{
     \includegraphics[width=0.70\textwidth]{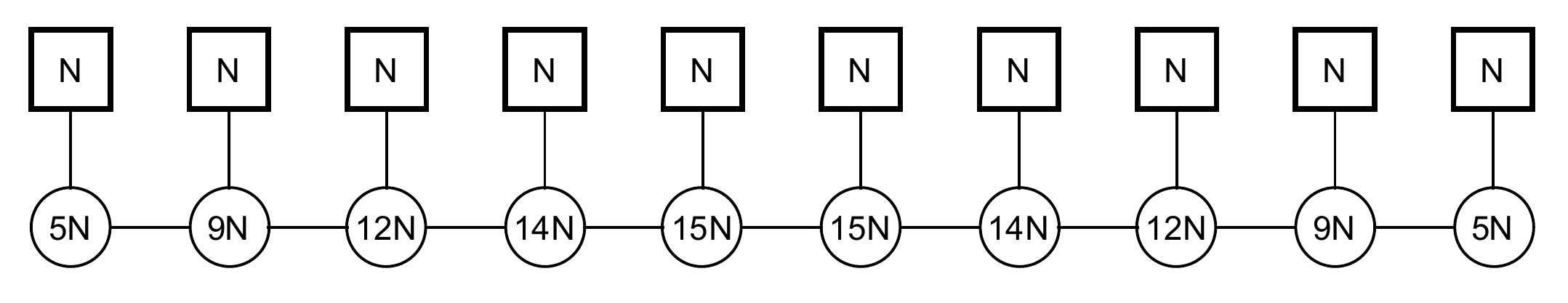}}\hspace{0.05\textwidth}\\
 \subfloat[\small ]{
    \includegraphics[width=0.60\textwidth]{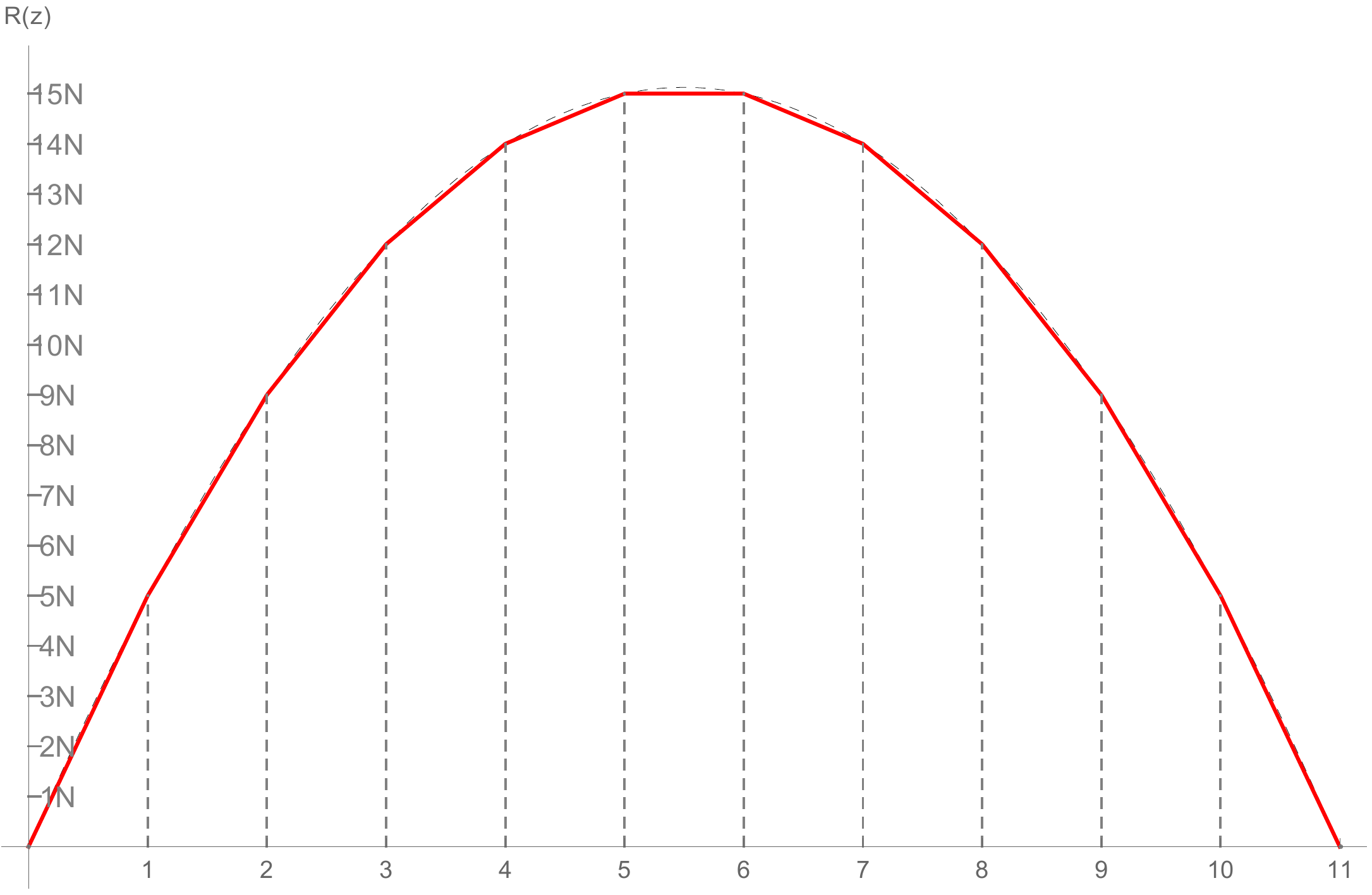}}   
\caption{(a) the quiver, and (b) the rank function (in red) for $n=5$ and $f=1$.}\label{fig_ParabolicQuiverTenNodes}
}
\end{figure}
The larger we make the ratio $n/f$, the closer the rank function of this quiver will approximate the function
\begin{equation}\label{eq_ContinuousRankParabolicQuiver}
R(z) = \left(n +\frac{f}{2}(1 - z)\right)z.
\end{equation}
See figure \ref{fig_ParabolicQuiverTenNodes}, where we illustrate this with an example for $n=5$ and $f=1$. 

The simplest way to think about this intuitively is that the flavour groups give the second derivative of the rank function (see section \ref{subsec_StartingFromFlavours}). When the flavour groups are constant everywhere, the rank function thus needs to have a constant second derivative and will be parabolic. As we choose the value of $n/f$ larger we will approximate a parabola with more points, which becomes continuous in the limit where we take $n/f \to \infty$.

We will now calculate the number of branes and central charges from the continuous rank function and show they indeed match as we take the limit $n/f \to \infty$. The number of branes can simply be calculated as
\begin{eqnarray}
& N_{D_p} =& \int dz\;R(z) = n\left( \frac{2n^2}{3f^2} + \frac{n}{f}\right), \label{eq_IntegralD6Branes}\\
& N_{D_{p+2}} =& -\int dz\;R''(z) = 2n. \label{eq_IntegralD8Branes}
\end{eqnarray}
Note this indeed agrees with the large $n/f$ limit of eq.(\ref{eq_NDd_DiscreteParabolicQuiver}) and (\ref{eq_NDdp2_DiscreteParabolicQuiver}).\\
\\
The central charge for the 4d 
-Maldacena case can be calculated as
\begin{equation}
c = \frac{2}{\pi^4} \int_0^{P+1}\lambda^2(z)\;dz=   n^2 \left( \frac{2n^2}{3f^2} + \frac{n}{f} \right)^2\\
\end{equation}
using that $\lambda(z) = R(z)$. Note this indeed corresponds with the large $n/f$ limit of eq.(\ref{eq_ccMG_ParabolicQuiver}). 

To calculate the central charge for the 6d Cremonesi-Tomasiello case we need the function $\alpha(z)$ that is proportional to the second integral of the rank function. Demanding that $\alpha'(z) = 0$ for $z = \frac12(\frac{2n}{f}+1)$, and $\alpha(z) = 0$ for $z = (\frac{2n}{f}+1)$ we find
\begin{equation}
\alpha'(z) = \frac{1}{12}\left( 6n + f(3-2z) \right)z^2 - \frac{(f+2n)^3}{24f^2},
\end{equation}
\begin{equation}
\alpha(z) = \frac{z^3}{24}\left( 2f + 4n - fz \right) - \frac{(f+2n)^3}{24f^2}z,
\end{equation}
which gives us for the central charge
\begin{equation}\label{eq_6dccContinuousParabolicQuiver}
c_{d=6} = - n^2 \left( \frac{34 n^5}{315 f^5}+\frac{17 n^4}{45 f^4}+\frac{17 n^4}{30 f^3}+\frac{17 n^2}{36 f^2}+\frac{17 n}{72 f}+\frac{17 f
   n}{1440}+\frac{17 f^2}{20160 n^2}+\frac{17}{240} \right).
\end{equation}
Note this result again corresponds with the large $n/f$ limit of eq.(\ref{eq_ccCT_ParabolicQuiver}).\\
\\
We thus find that in the limit that $n/f \to \infty$ the continuous rank function from eq.(\ref{eq_ContinuousRankParabolicQuiver}) indeed captures the correct number of D-branes and central charges. 

\subsection{Sinusoidal Quiver}\label{subsec_SinusoidalQuiver}
We will now show another example of a continuous rank function of the form
\begin{equation}
R(z) = A \sin \omega z.
\end{equation}
This particular example is especially interesting as it was recently shown in \cite{Filippas:2019puw} that the bosonic sector of the string is integrable on the AdS$_7$ Cremonesi-Tomasiello geometry corresponding to this rank function.

One can think of this rank function as a Fourier mode. In this context the Gaiotto-Maldacena AdS$_5$ solution corresponding to this rank function was discussed in detail in \cite{Aharony:2012tz} (section 4), its analytic continuation has been discussed in \cite{Lin:2005nh} (see eq.(2.44) of that paper).\\
\\
For this sinusoidal quiver, the rank function has the property $R(z) = - \omega^{-2} R''(z)$. This means all the flavour groups are proportional to the colour groups, as the number of flavour branes everywhere has to be proportional to the second derivative of the rank function. We can, therefore, approximate this continuous rank function by a discrete one, using as a defining condition that the number of flavours is everywhere proportional to the colour groups.\footnote{Note that the shortest possible quiver of this type is the $N_F = 2N_C$ quiver, when we set $\epsilon = \frac12$, 
}
\begin{equation}\label{eq:proportionalityConstant}
N_k = \epsilon\;F_k.
\end{equation}
If we take this to be the defining condition for the quiver and set $N_0 = F_0 = 0$ at the beginning of the interval, we can use the consistency condition in eq.(\ref{eq_AnomalyCancelationCondition}) to recursively define the values of $N_k$ for a quiver or arbitrary length from the consistency condition
\begin{eqnarray}
&N_{k} &= 2N_{k-1} - F_{k-1} - N_{k-2},\\
&        &= \left( 2 - \frac{1}{\epsilon} \right) N_{k-1} - N_{k-2}.
\end{eqnarray}
For this example it is more complicated than for the parabolic quiver to give a closed expression for value of the different ranks $N_k$
\begin{equation}\label{eq_GeneralNSinusoidalQuiver}
N_k = nk - \sum_{i=1}^{k-1} f_k,
\hspace{1cm}
\text{where}
\hspace{1cm}
F_{k-1} = \frac{1}{\epsilon}\left( n(k-1) - \sum_{i=1}^{k-2} f_k \right).
\end{equation}
As the rank $N_k$ is given in terms of the previous ranks and flavours, and the number of flavours is now proportional to the rank - which in turn is given in terms of all the previous ranks and flavours - we find that $N_k$ is given as a recursive summation.

\begin{figure}[t]
{
 \centering
 \subfloat[\small ]{
   \label{fig_sinusoidalQuiver3NodesA}
     \includegraphics[width=0.45\textwidth]{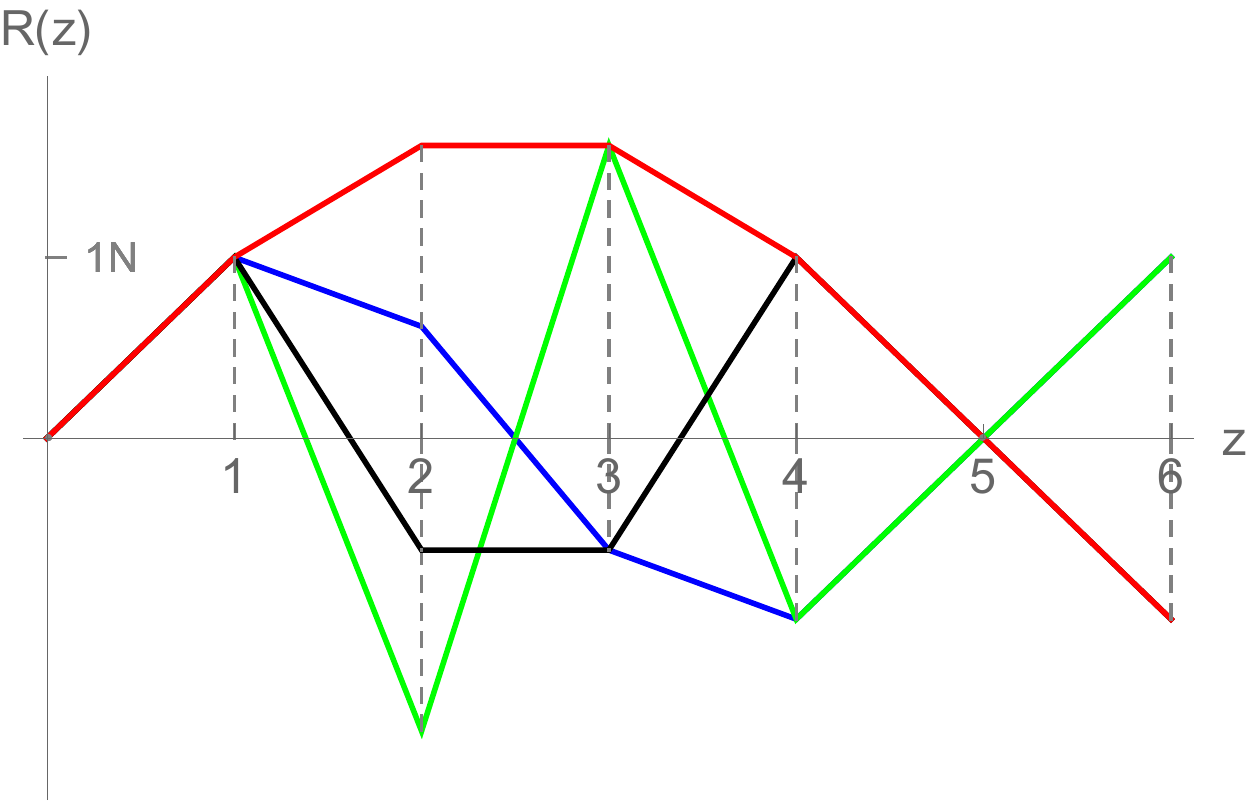}}\hspace{0.05\textwidth}
 \subfloat[\small ]{
   \label{fig_sinusoidalQuiver3NodesB}
     \includegraphics[width=0.45\textwidth]{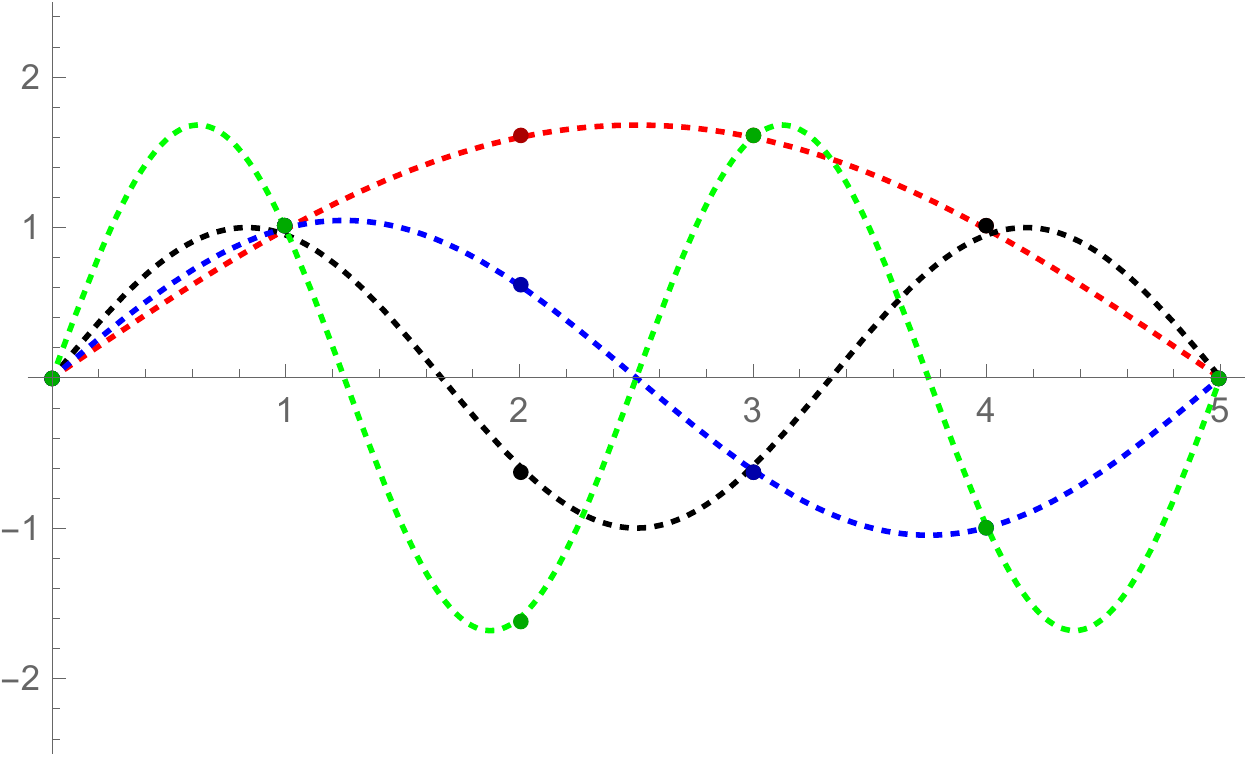}}\hspace{0.05\textwidth}\\
 \subfloat[\small ]{
   \label{fig_sinusoidalQuiver3NodesC}
    \includegraphics[width=0.45\textwidth]{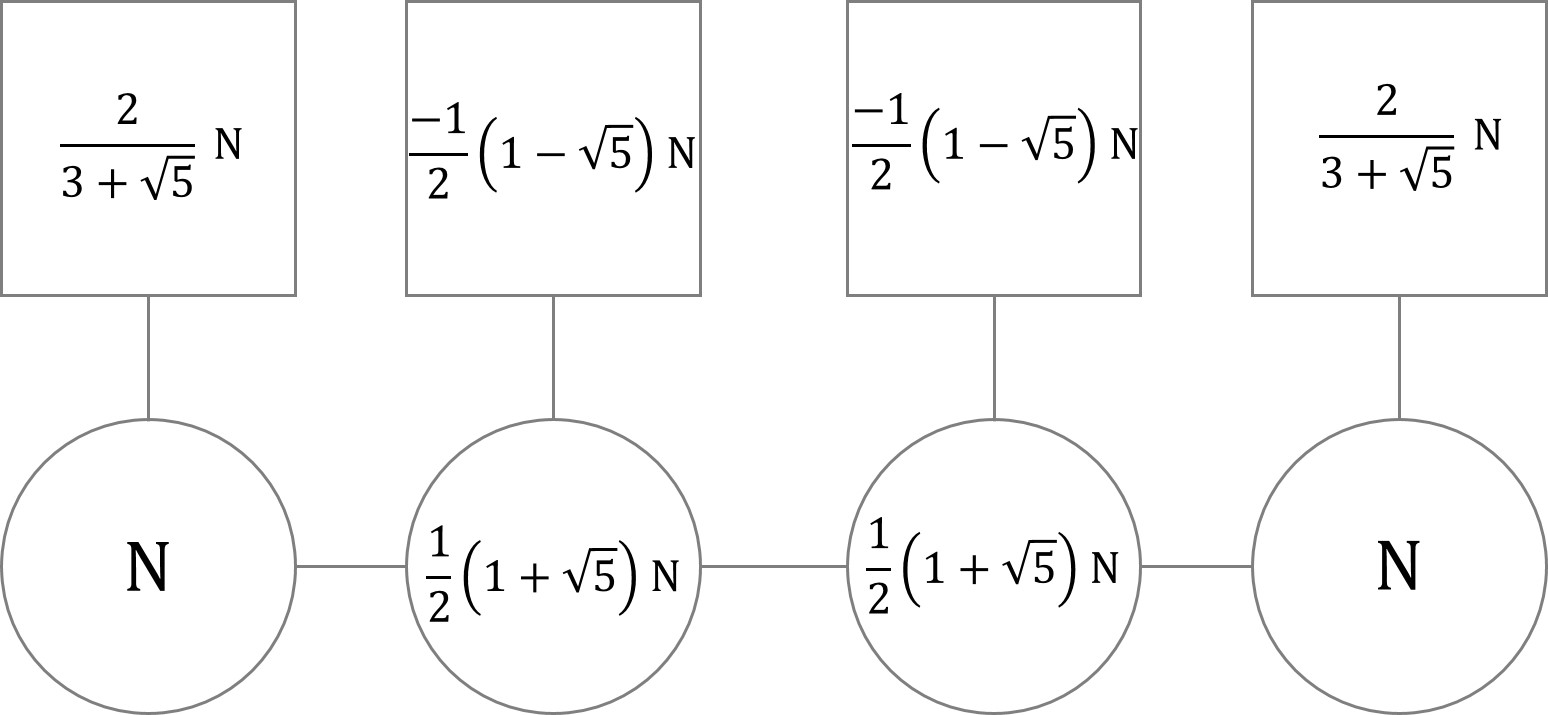}}   
\caption{(a) Piece wise continuous rank functions for the different values of $\epsilon$: $\epsilon^1_+$ (red), $\epsilon^1_-$ (black), $\epsilon^2_+$ (green) and $\epsilon^2_-$ (blue). (b) Corresponding continuous rank functions that are approximated by these discrete ones. (c) Quiver diagram for $\epsilon^1_+$. Note that the values of all flavour groups are proportional to $2 / (3 + \sqrt{5})$ times the corresponding colour groups.}\label{fig_sinusoidalQuiver3Nodes}
}
\end{figure}

If we would work this out and set $N_1 = n$ we find first the terms look like
\begin{eqnarray}
& N_2 &= 2n - \frac{n}{\epsilon} \\
& N_3 &= 3n -\frac{4 n}{\epsilon } + \frac{n}{\epsilon ^2},\\
& N_4 &= 4n -\frac{10 n}{\epsilon }+\frac{6 n}{\epsilon ^2}-\frac{n}{\epsilon ^3},\\
& N_5 &= 5n-\frac{20 n}{\epsilon }+\frac{21 n}{\epsilon
   ^2}-\frac{8 n}{\epsilon ^3}+\frac{n}{\epsilon ^4}, \\
& N_6 &= 6n -\frac{35 n}{\epsilon }+\frac{56 n}{\epsilon ^2}-\frac{36 n}{\epsilon
   ^3}+\frac{10 n}{\epsilon ^4}-\frac{n}{\epsilon ^5},\\
& \hdots &
\end{eqnarray}
As an example let us now obtain a quiver with only three colour groups that approximates a sinusoidal quiver. This means we want $N_5 = 0$, so we can solve for $\epsilon$
\begin{equation}
5n \epsilon^4 - 20 n \epsilon^3 + 21n \epsilon^2 - 8n \epsilon + n = 0,
\end{equation}
which has the roots
\begin{equation}
\epsilon_\pm^1 = \frac12( 3 \pm \sqrt{5} )
\hspace{1cm}
\text{and}
\hspace{1cm}
\epsilon^2_\pm = \frac{1}{10}(5 \pm \sqrt{5}).
\end{equation}
The different roots will now all correspond to quivers with rank functions that are piece wise continuous approximations of sine functions (see figure \ref{fig_sinusoidalQuiver3Nodes}), all vanishing at $z = N_5$.
We can generalise this example to generate piecewise continuous approximations to the sinusoidal rank function of any length, by requiring that the appropriate $N_k = 0$. Eq. (\ref{eq_GeneralNSinusoidalQuiver}) will then give a $(k-1)$-th order polynomial in $\epsilon$, the different roots of which will correspond to piecewise continuous approximations to sinusoidal functions of different wavelengths all vanishing at $z = k$. The largest of these roots will always correspond to an approximation of a sine with period $2k$. 

As we take in eq.(\ref{eq:proportionalityConstant}) the 't Hooft limit $\epsilon = \frac{N_n}{F_n} \to \infty$, the length of the largest sine wave becomes infinite. In this limit, the piecewise continuous rank functions will have the same numbers of colour and flavour branes associated with it as one would obtain from the continuous rank function, the way we illustrated earlier for the parabolic quiver.

\subsubsection*{Integer Numbers of Branes}
As can be seen in figure \ref{fig_sinusoidalQuiver3Nodes}, for finite $\epsilon$ the different roots will, in general, be non-integer numbers. Since $N_k = \epsilon F_k$, the number of flavours is then a non-integer number times the number of colours. One might be troubled that this is inconsistent when we think of these rank function as originating from brane-set ups where we require there to be an integer number of branes.

\begin{figure}[]
{
 \centering
 \subfloat[\small ]{
     \includegraphics[width=0.3\textwidth]{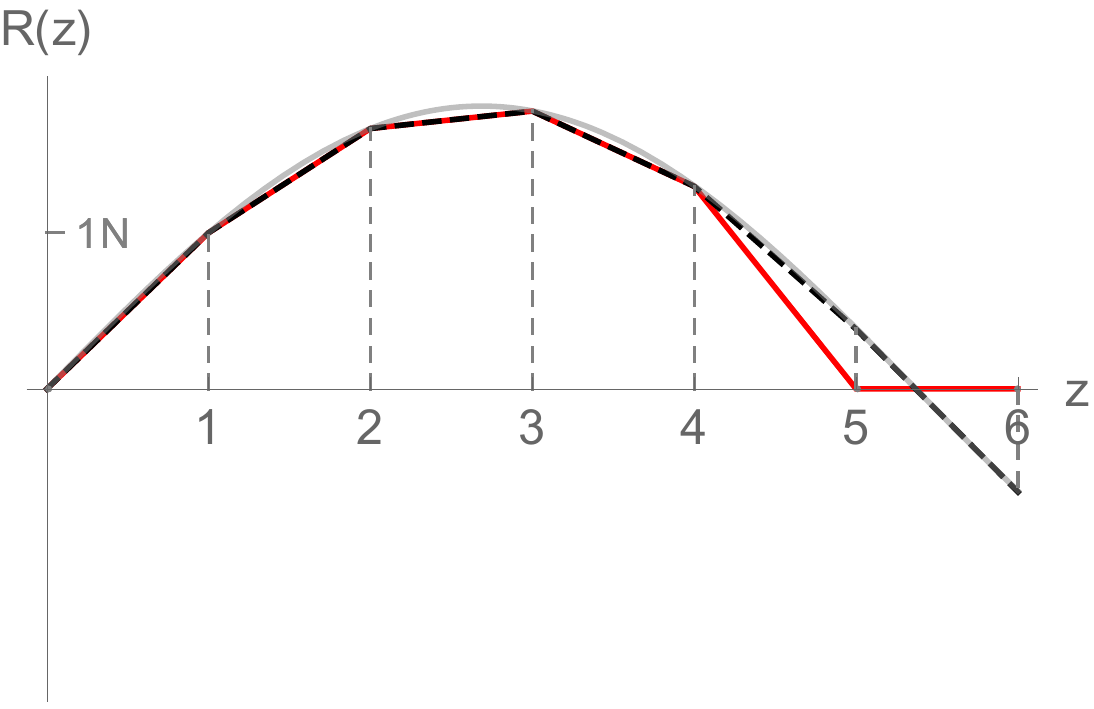}}\hspace{0.05\textwidth}
 \subfloat[\small ]{
    \includegraphics[width=0.2\textwidth]{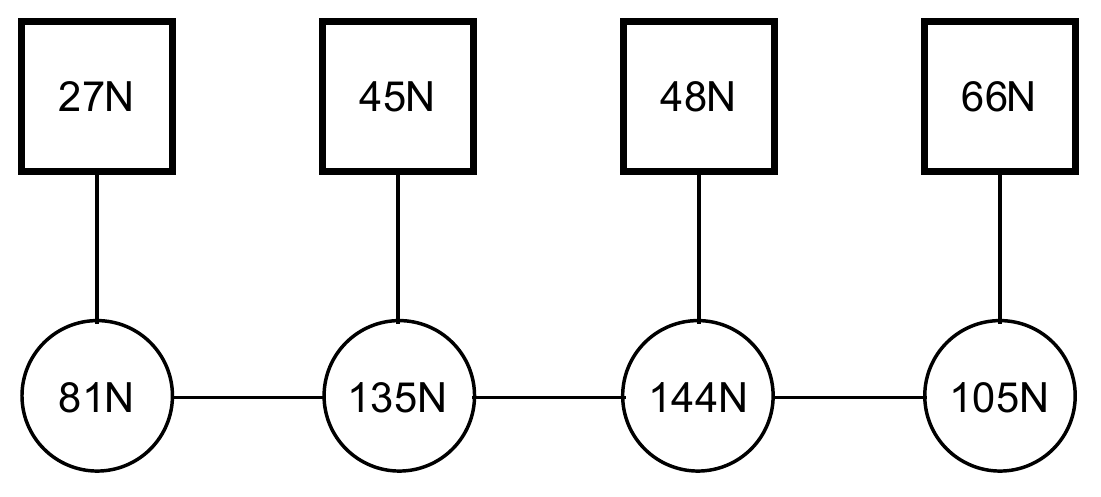}}   
\caption{\textbf{(a)} The rank function for $\epsilon = 3$ (black) and the manually adjusted rank function (red) where $N_5 = 0$. \textbf{(b)} The correspoding quiver diagram for the `ammended'  rank function (red), after rescaling.}\label{fig_truncateQuivers}
}
\end{figure}

Let us first remark that for the continuous rank function where $\epsilon = N_n /F_n \to \infty$ this is not a problem. We will now address this problem for finite sinusoidal rank functions and discuss a how one can amend them to obtain quivers where the numbers of colour and flavour branes is integer.\\
\\
As an example, we will take $\epsilon = 3$ so that $N_k = 3 \;F_k$. When we choose an arbitrary integer (or rational) value for $\epsilon$ (as we do here) it is not guaranteed that the rank function will exactly go to zero at some point. For his example we obtain a rank function where $N_5 > 0$ and $N_6 < 0$. To get a rank function of five nodes we add some additional flavours to $F_4$ by hand, and force $N_5 = 0$, see figure \ref{fig_truncateQuivers}. The resulting ranks of the flavour groups are then
\begin{equation}
\Big\{ 0,\; 1,\; \frac53,\; \frac{16}{9},\; \frac{35}{27},\; \frac{31}{81},\; -\frac{160}{243} \Big\}
\hspace{1cm}\to \hspace{1cm}
\Big\{ 0,\; 1,\; \frac53,\; \frac{16}{9},\; \frac{35}{27},\; 0 \Big\}
\end{equation}
Note that for an integer value of $\epsilon$ the $N_n$ are fractions with $\epsilon^{n}$ in the denominator. To obtain a rank function of length $n$, corresponding to a brane set-up with integer numbers of colour and flavour branes we rescale using eq.(\ref{eq_SugraRescaling}) with $N = \epsilon^n$. The result then corresponds to the quiver diagram in fig. \ref{fig_truncateQuivers}. Note that the resulting quiver has for all nodes $N_k = \epsilon F_k$, except the last one as we changed it by hand to truncate the quiver consistently. 
This procedure thus allows us to create consisted rank functions that approximate sinusoidal ones everywhere except at its final value. 

As we take the limit $\epsilon \to \infty$ the sinusoidal rank function becomes infinitely long and this error becomes again negligible.

\section{Smeared Branes}\label{sec_SmearedBranes}
In the previous section, we started by defining infinitely long quivers and argued they are described by continuous rank functions in their gravitational dual. We will now reason the other way around: we start by considering a supergravity solution with a continuous rank function, and we show this implies the flavour branes that backreact on the geometry are smeared. We then show the correct way to interpret the corresponding brane set-up is as an infinitely long quiver described in the previous section.\\
\\
When we consider the supergravity solutions in eqs.(\ref{eq:10dGaiottoMaldacenaNS}) and (\ref{eq:TomasielloGeometryGeneral}) with a rank function $R(z)$, the second derivative $R''(z)$ indicates the positions of the flavour branes backreacting on the geometry. 

When we consider a piecewise linear rank function $R(z) \in C^{(0)}$ (as in section \ref{subsec_ScalinginiteQuiver}), the positions of the flavour branes are given by delta functions $R''(z) = \sum_i \delta(z_i)$. When we consider a continuous rank function $R(z) \in C^{(\infty)}$, the second derivative will no longer be a series of delta functions, but will instead be continuous $R''(z) = \rho(z)$. This implies the flavour branes are no longer located at fixed points, but are instead smeared along the $z$-direction of the geometry. The geometry is now a solution of the equations of motion for the action
\begin{equation}
S_{Type\;II} +  \rho(z)\;S_{BIWZ},
\end{equation}
which differs from the action in eq.(\ref{eq_TypeIIBIWZAction}) in that the $D_{p+2}$-flavour branes are not located at fixed points given by delta functions, but instead are smeared with a certain density $\rho(z)$. The backreaction of smeared flavour branes was first introduced in \cite{Bigazzi:2005md}, and further studied in a 10d context in \cite{Casero:2006pt, Nunez:2010sf}
\begin{figure}[h!]
\centering
\subfloat[]{\label{fig_flavourDensityDiscrete}
     \includegraphics[width=0.3\textwidth]{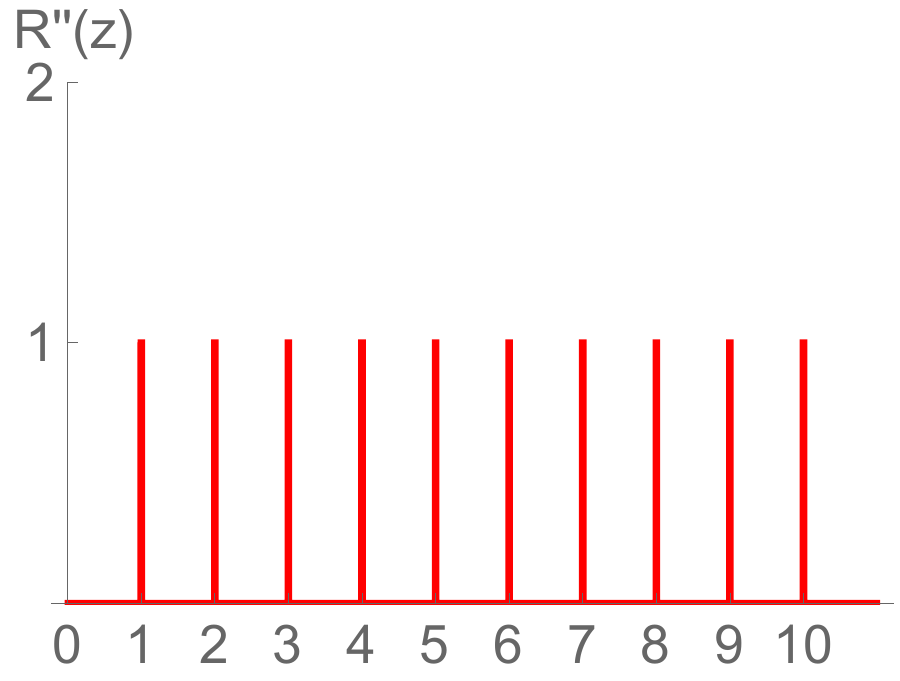}}
\subfloat[]{\label{fig_flavourDensityContinuous}
    \includegraphics[width=0.3\textwidth]{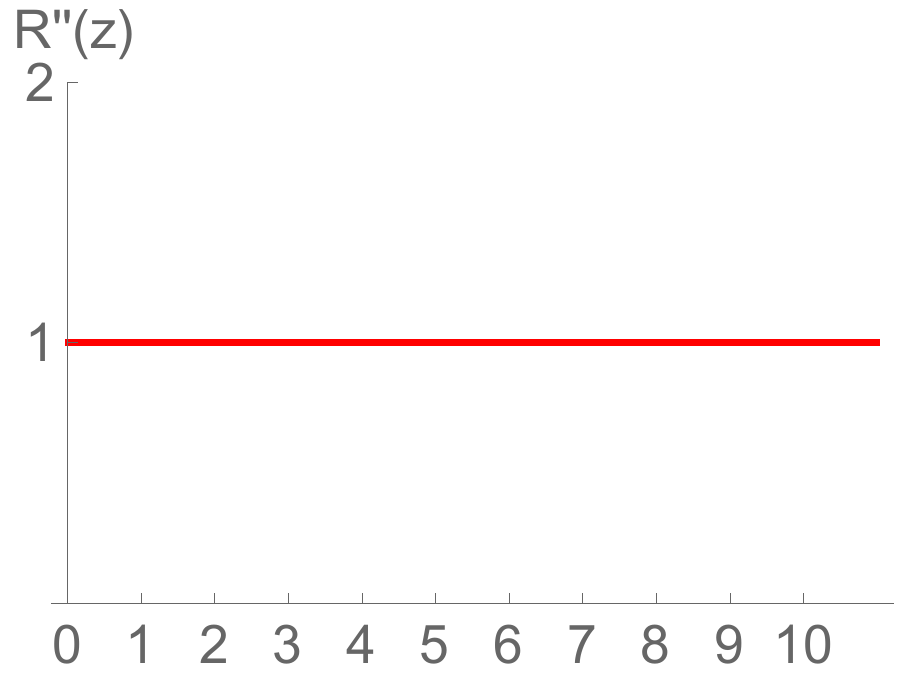}} 
\caption{The positions of the flavour branes for (a) the discrete and (b) the continuous parabolic rank function of figure \ref{fig_ParabolicQuiverTenNodes}. The continuous rank function implies a continuously smeared distribution of flavour branes.
} \label{fig_flavourDensities}
\end{figure}\\
\\
Since the $D_{p+2}$ flavour-branes act as sources for the $F_{8-(p+2)}$ RR-flux, the Bianchi identities are violated at the points where these flavour branes are located. In addition, the curvature will typically diverge at the points where these flavour branes are located.

When we consider a continuous rank function, the delta functions are `smeared' and the resulting geometries no longer has singular points (as can be verified by inspecting the dilaton and curvature). 
This smearing then causes the Bianchi identities to be violated everywhere along the $z$-direction\footnote{See Appendix A of \cite{Nunez:2010sf} for an interesting relation between the violation of the Bianchi identities, supersymmetry, and calibrated geometry.}
\begin{equation}
dF_{8-(p+2)} = \rho(z).
\end{equation}
Unlike the examples studied in \cite{Nunez:2010sf}, we can not just add a smeared number of flavour branes to the half-BPS backgrounds we consider in this paper. Since the numbers of colour and flavour branes are not independent of each other, but are both related to the rank function $R(z)$ though the consistency condition, eq.(\ref{eq_CFConsistencyCondition}), smearing the flavour branes then implies the number of colour branes has to become infinite as well. This can also be seen from the Page charges for the total number of flavour and colour branes in eq.(\ref{GMcargasgood}) and (\ref{cargasgood}), which are both related to the rank function as
\begin{eqnarray}
& N_{D_{p+1}} &= \int_0^{P+1}dz\;R''(z), \\
& N_{D_p} &= \int_0^{P+1}dz\;R(z).
\end{eqnarray}
Since the colour branes have to end on NS5-branes in the corresponding Hanany-Witten set-up, we cannot have let the number of colour branes go to infinity without simultaneously letting the number of NS5-branes go to infinity. This ensures that all the colour branes end on NS5-branes.

When we consider a finite continuous rank function $R(z)$ for $z \in[0,1]$, increasing the number of NS5-branes is done by the scaling $z \to \omega z$ we discussed earlier in section \ref{subsec_ScalinginiteQuiver}. This means we have to rescale the continuous rank function to become infinitely long. We thus find that we can interpret any (concave) continuous rank function after `stretching' it as accurately approximated by a piecewise continuous quiver of infinitely many points (for which there is a well-defined brane set-up), as we showed in the previous section.

We can think of these continuous rank functions with smeared branes as `macroscopic' descriptions of the infinite quivers defined in the previous section, where the infinite numbers of branes become continuous distributions. 

\subsection{Orthogonal Bases of Quivers}\label{sec_OrthogonalBases}
Since the rank functions can be decomposed into various orthogonal bases of functions, this naturally brings about the question if such bases have some interpretation in the dual SCFTs. 

The differential equations for the functions that specify the half-BPS AdS geometries are linear (see eqs. (\ref{eq:AlphaThird}) and (\ref{eq:GaiottoMaldacenaLaplaceEquation}) for the AdS$_5$ and AdS$_7$ cases respectively). We can therefore obtain any solution as a superposition of modes. Since we have shown in the previous sections how to interpret continuous rank functions as corresponding to infinite quivers, we now try to find a field theory interpretation for this basis of functions
\begin{equation}\label{eq_fundamentalModesQuivers}
\alpha_{\text{AdS}_7}(z) = -\sum_{n=1}^\infty A_n \sin (\omega_n z), \hspace{1cm}
V_{\text{AdS}_5}(\sigma, z) = - \sum_{n=1}^\infty c_n K_0 (\omega_n \sigma) \sin(\omega_n z).
\end{equation}
The rank functions corresponding to these solutions will be sinusoidal quiver with the ranks of its colour groups proportional to its flavour groups $N_n = \epsilon F_n$ that we introduced in section \ref{subsec_SinusoidalQuiver}. It would be very interesting to better understand if this linearity of the differential equations (\ref{eq:AlphaThird}) and (\ref{eq:GaiottoMaldacenaLaplaceEquation}) reflects some deeper property of the SCFTs.\\
\\
As an example, we cite the result that the central charge in eq.(\ref{eq_MGCentralCharge}) and eq.(\ref{eq_TCCentralCharge}) of an arbitrary 4d or 6d quiver SCFT, is directly given as a superposition of the central charges of these modes \cite{Nunez:2018ags, Nunez:2019gbg}\footnote{See section 2.3 of \cite{Nunez:2018ags}, section 5 of \cite{Filippas:2019puw}, see also \cite{Nunez:2019gbg}.}
\begin{equation}
c_{6d} = \frac{-2^8}{3^8\times16\times G_N} \int_0^{P+1}dz\;\alpha''(z) \alpha(z),
\hspace{2cm}
c_{4d} = \frac{2}{\pi^4}\int_0^{P+1}dz\;\lambda^2(z).
\end{equation}
Note that since the different modes are orthogonal, and $\alpha''(z) \sim \alpha(z)$, the above computation becomes the sum over the contributions to the central charge from each mode. It would be very interesting to better understand this relation from the SCFT point of view.

In general, we expect other observables of the SCFTs to not linearly depend on the modes in eq.(\ref{eq_fundamentalModesQuivers}). This can be seen as the AdS$_5$ and AdS$_7$ geometries in eqs.(\ref{eq:10dGaiottoMaldacenaNS}, \ref{eq:TomasielloGeometryGeneral}) depend on various combinations of the functions $\alpha(z)$ and $V(z, \sigma)$ and their derivatives. Even though the differential equations for $\alpha(z)$ and $V(z, \sigma)$ are linear, this linearity will not be reflected in the geometry, or the string dynamics on these spaces. As a result, observables like Wilson lines (that depend on the string dynamics) or the entanglement entropy of certain regions (which depends on the embedding of minimal surfaces) will not be linear in these modes.


\subsection*{Polynomial Bases}
It would be interesting to see if more quantities could be calculated directly from the rank function, and if other bases of functions - orthogonal under different inner products - might lend themselves to particular calculations.

Let us briefly remark that when constructing a basis of functions, we do not have to require the rank functions for these basis functions to satisfy the consistency condition. It is sufficient that the final superposition of modes satisfy this relation. 
\begin{equation}
F_n = 2N_n - N_{n+1} - N_{n-1}.
\end{equation}
One could use the tools of section \ref{subsec_ParabolicQuiver} to construct in an analogous fashion a basis of orthogonal polynomial rank functions.


%

\section{Half-BPS AdS$_7$, AdS$_5$ and AdS$_3$ Geometries }\label{sec_AdS5andAdS7}
We will list the various D$_p$-D$_{p+2}$-NS5-brane systems and their holographic duals that have been extensively studied in the literature. In all of these AdS geometries, there is one internal direction $z$, that geometrically encodes the quiver structure of the corresponding brane set-up. All of the warp factors in these AdS geometries are completely determined by one or more functions $\mathcal{F}(z, \ldots)$ that always depends on $z$ and possibly on an additional direction of the internal space $\Sigma_2$. In order to satisfy the BPS equations, this function $\mathcal{F}$ has to satisfy a linear differential equation, where the rank function $R(z)$ is related to the boundary conditions of $\mathcal{F}$.
For general $p$, the resulting geometries are of the form shown in table \ref{table_AdSGeometries}. 
\begin{table}[h]
\centering
\begin{tabular}{l|l}
\textbf{p} & \textbf{AdS$_{p+1}$}                        \\ \hline
6          & $\left( AdS_7\times S^2 \right) \ltimes \Sigma$     \\ \hline
5          & $\left( AdS_6 \times S^2 \right) \ltimes \Sigma_2$           \\ \hline
4          & $\left( AdS_5\times S^2 \times S^1 \right) \ltimes \Sigma_2$ \\ \hline
3          & $\left( AdS_4\times S^2 \times S^2 \right) \ltimes \Sigma_2$ \\ 
\end{tabular}\caption{Near horizon geometries for the different $\frac12$BPS $D_p$-$D_{p+2}$-NS5-brane systems.}\label{table_AdSGeometries}
\end{table}\\
\\
In this section we will discuss the cases $p = 6$ and $p=4$, as our ideas apply most directly to these cases. For these AdS$_7$ and AdS$_5$ geometries, $R(z)$ is directly related to (the boundary values of) the derivatives of the functions $\mathcal{F}$. We show explicitly in both cases how one can define the supergravity solutions corresponding to continuous rank functions.\\
\\
The cases $p=3$ and $p=5$ are a bit different. The quiver structure of the dual field theories for these AdS$_4$ and AdS$_6$ geometries is encoded in the poles of the functions $\mathcal{F}$ at the boundary of $\Sigma_2$. This makes it not directly obvious how we could define the analogue of a continuous rank function for those cases. 

In addition, the consistency condition that related the ranks of the flavour groups to the second derivative of colour rank function is no longer valid in these cases. Our approach to construct infinite quivers in section \ref{subsec_StartingFromFlavours} by defining the flavour groups does therefore not directly apply here. 

In Appendix \ref{App_AdSGeometries} we discuss these cases, and make some observations on infinite linear rank functions that were interpreted as involving smeared branes \cite{Lozano:2016wrs, Lozano:2018pcp}.

\subsection{AdS$_7$}\label{subsec_AdS7}
D6-D8-NS5 brane set-ups give rise to 6d $\mathcal{N} = (1,0)$ supersymmetric QFTs \cite{Hanany:1997gh}. For conformal field theories the symmetry group of the worldvolume theory is enhanced to $OSp(2,6|2)$. These SCFTs are expected to have a Massive Type IIA dual description \cite{Gaiotto:2014lca, Cremonesi:2015bld, Bobev:2016phc} in terms of a geometry of the form
\begin{equation}
\left( AdS_7\times S^2 \right) \ltimes \mathbb{R},
\end{equation}
where the metric and field content are given by
\begin{eqnarray}\label{eq:TomasielloGeometryGeneral}
& & ds^2 = \sqrt{2}\pi \left(-\frac{\alpha}{{\alpha''}} \right)^{1/2} \left[ 8\;ds^2_{AdS_7} + \frac{\alpha'' \alpha}{\Delta}\;d \Omega_2^2 +
\frac{\alpha''}{\alpha} dz^2 \right],\nonumber\\
& & B_2=\pi \left(\frac{\alpha {\alpha'}}{\Delta} - z\right) d\Omega_2,\qquad  e^{\phi}= 3^4 \left( (2\pi^2)^5 \frac{(-\alpha/ {\alpha''})^3}{\Delta^2} \right)^{1/4},\\
& & F_2= \left( \frac{{\alpha''}}{162 \pi^2}+ \frac{\pi F_0 \alpha {\alpha'}}{\Delta}  \right)d\Omega_2, \qquad  F_0= F_0(z).\nonumber
\end{eqnarray}
Here $\Delta = {\alpha'}^2-2 \alpha {\alpha''}$, and we defined $d\Omega_2^2 = d\chi + \sin\chi^2 d\xi$ and $d\Omega_2 = \sin\chi\;d\chi\wedge d\xi$. The entire geometry is determined by the function $\alpha(z)$, which has to vanish at the end of the $z$-interval, and satisfy the differential equation 
\begin{equation}\label{eq:AlphaThird}
{\alpha'''}(z)=-162 \pi^3 F_0.
\end{equation}
This function $\alpha(z)$ only depends on the $z$-direction, and is directly related to the rank function of the corresponding brane set-up as
\begin{equation}\label{TomasielloGeometryRelationToRankFunction}
R(z) = \frac{-1}{81 \pi^2} \alpha''(z).
\end{equation}
When there is a finite number of flavour branes, $F_0$ is constant and discontinuous, and the corresponding function $\alpha(z) \in C^{(2)}$ is a piece-wise continuous third order polynomial in $z$. See section 2 of \cite{Filippas:2019puw} for a more detailed review. 

When we consider a continuous rank function $R(z) \in C^{(\infty)}$ one can directly obtain the corresponding function $\alpha(z)$ from eq.(\ref{TomasielloGeometryRelationToRankFunction}).

\subsection{AdS$_5$}\label{subsec_AdS5}
D4-D6-NS5 brane set-ups give rise to 4d $\mathcal{N} = 2$ supersymmetric QFTs, like e.g. Seiberg-Witten theory \cite{Seiberg:1994rs} and Gaiotto theories that can also be obtained from the reduction of a single M5-brane wrapping a punctured Riemann surface \cite{Gaiotto:2009we}. At the conformal fixed point, the symmetry group is enhanced to $SU(2,2|2)$ and these set-ups are thought to have a near-horizon geometry \cite{Gaiotto:2009gz} of the form
\begin{equation}
\left( AdS_5\times S^2 \times S^1 \right) \ltimes \Sigma_2,
\end{equation}
with $\Sigma_2(z, \sigma)$ a two-dimensional Riemann surface. The NS-NS and RR-RR sectors are given by
\begin{eqnarray}\label{eq:10dGaiottoMaldacenaNS}
& ds^2 =& \left( \frac{2\dot{V} - \ddot{V}}{V''} \right)^{1/2} \left[ 4\;ds^2_{AdS_5} + \frac{2 V'' \dot{V}}{\Delta} d\Omega^2_2 + \frac{2V''}{\dot{V}} (d\sigma^2 + dz^2 ) + \frac{4V''\sigma^2}{2\dot{V} - \ddot{V}}d\beta^2 \right], \nonumber\\
& B_2 =& 2\left( \frac{\dot{V}\dot{V}'}{\Delta} - z \right)\;d \Omega_2, \hspace{2cm}
  e^{\phi} = \left( 4\frac{(2\dot{V} - \ddot{V})^3}{V'' \dot{V}^2 \Delta^2} \right)^{1/4}.\\
& A_1 =& 2 \frac{2 \dot{V} \dot{V}'}{2\dot{V} - \ddot{V}}\;d\beta, \hspace{3.2cm} 
  C_3 = -4 \frac{\dot{V}^2 V''}{\Delta}\;d\beta\wedge d\Omega_2 .\nonumber
\end{eqnarray}
Note the entire geometry here is determined by the function $V(z, \sigma)$ and its derivatives. Here $\Delta = (2\dot{V} - \ddot{V} )V'' + (\dot{V}')^2$, and the dots and primes indicate the derivatives $\dot{V} = \sigma \partial_\sigma V$, and $V'= \partial_z V$. The potential $V(z, \sigma)$ has to satisfy the Laplace equation
\begin{equation}\label{eq:GaiottoMaldacenaLaplaceEquation}
\ddot{V} + \sigma^2 V'' = 0,
\end{equation}
supplemented by the boundary conditions
\begin{equation}
\sigma \partial_\sigma V \Big|_{\sigma=0} = \lambda(z), \hspace{2cm} V\Big|_{\sigma\to\infty} = 0.
\end{equation}
These boundary conditions are now directly related to the rank function of the quiver structure in the corresponding brane set-up as
\begin{equation}\label{eq_GMRank}
\lambda(z) = R(z),
\end{equation}
and has to vanish at the beginning and end of the z-interval. See \cite{ReidEdwards:2010qs, Aharony:2012tz, Nunez:2019gbg} for reviews. 
For a piece-wise continuous rank function, the solution for $V(z, \sigma)$ can be constructed as a superposition of Maldacena-N\'{u}\~{n}ez solutions.
\begin{eqnarray}
& V_{MN} =& \frac12\sqrt{\sigma^2 + (N+z)^2 } - \frac12 (N+z)\sinh^{-1} \left( \frac{N+z}{\sigma} \right) \\
& & \hspace{5cm} - \frac12\sqrt{\sigma^2 + (N-z)^2 } + \frac12 (N-z)\sinh^{-1} \left( \frac{N-z}{\sigma}. \right) \nonumber
\end{eqnarray}
for which
\begin{equation}
\dot{V}_{MN} = \frac12 \sqrt{\sigma^2 + (N+z)^2 } - \frac12 \sqrt{\sigma^2 + (N-z)^2 }.
\end{equation}
so that they correspond to the a rank function
\begin{equation}
\lambda_{MN}(z) = \frac12 |N+z| - \frac12 |N-z| = \begin{cases}
  z & \text{if} \quad 0 \leq z \leq 1\\      
  N & \text{if} \quad z \geq 1
\end{cases}
\end{equation}
One can thus construct the function $V(z, \sigma)$ for a piece wise linear rank function $R(z) \in C^{(2)}$ as a sum of these Maldacena-N\'{u}\~{n}ez solutions. To construct the function $V(z, \sigma)$ for a continuous rank function $R(z) \in C^{(\infty)}$ one could take an increasing number of these solutions to approximate the rank function with an increasing number of points.\\
\\
An alternative way to define a solution for $V(z, \sigma)$ for a continuous rank function $R(z) = \lambda(z) \in C^{(\infty)}$ is to write a general solution as a series
\begin{equation}
V_{\omega_n}(\sigma, z) = - \sum_{n=1}^\infty c_n K_0 (\omega_n \sigma) \sin(\omega_n z), \hspace{2cm} \omega_n = \frac{n\pi}{N_5}, \label{eq:VpotentialFourierExpansion}
\end{equation}
Note the boundary profiles of $\dot{V}(z, \sigma)\Big|_{\sigma=0}$ for these modes are sinusoidal. As we explained in section \ref{sec_OrthogonalBases}. Each of these modes can be thus thought of as an infinite quiver where $N_n = \epsilon F_n$. Since we know the profile of these basis functions along the $\sigma$-direction we can construct any continous rank function as a superposition of them.
\begin{figure}[t]
{
\center
   \includegraphics[width=0.45\textwidth]{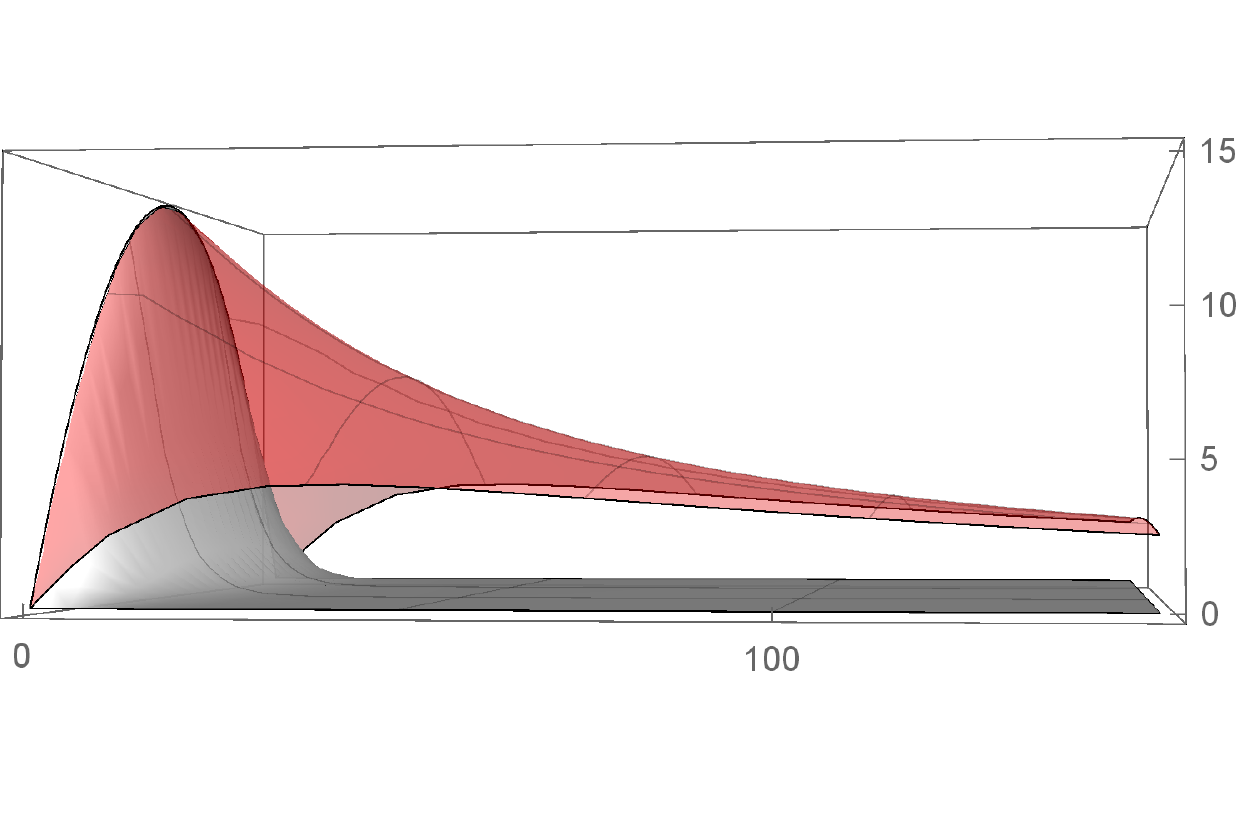}
      \includegraphics[width=0.45\textwidth]{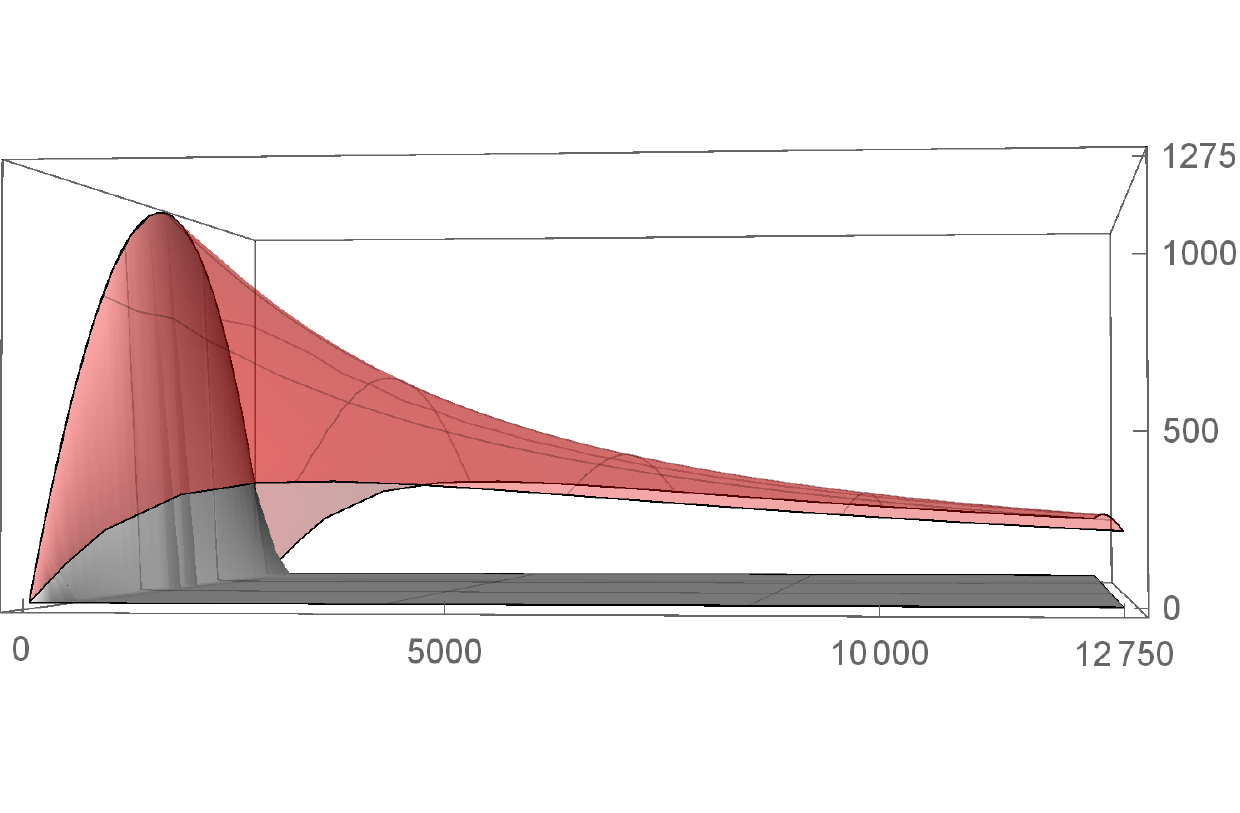}
\caption{The functions $\dot{V}$ for the Gaiotto-Maldacena solution corresponding to the rank functions in figure \ref{fig_ParabolicQuiverTenNodes}, \textbf{(red)} Piece-wise continuous solution $R(z) \in C^{(2)}$ at $\sigma = 0$, composed as a superposition of Maldacena-Nunez solutions. \textbf{(grey)} Continuous solution $R(z) \in C^{(\infty)}$ at the boundary, composed as a superposition of modes. \\
 We plot two examples for the parabolic quiver: \textbf{(left)} $n=5$ and $f=1$ and 10 points in the discrete approximation. \textbf{(right)} $n=50$ and $f=1$ and 100 points in the discrete approximation. Note the retained difference along the $\sigma$-direction as the functions approach the same boundary profile.}\label{fig_ParabolicQuiverGaiottoMaldacena}
}
\end{figure}

We show an example of this for the piecewise, and continuous parabolic quivers in figure \ref{fig_ParabolicQuiverGaiottoMaldacena}. Note the continuous rank function has a corresponding function $\dot{V}(z, \sigma)$ that goes faster to zero than the piece-wise linear function that approximates it, even as we increase the number of points in the approximation.
Even though the functions approach the same boundary profile $\lambda(z)$, their discreteness or continuity influences their $\sigma$-dependence. We suggest the continuous function (grey) is the correct one to appropriately describe the infinite quiver.

\subsection{AdS$_3$}
Let us briefly point out the same logic also applies to the recently constructed class of quarter-BPS AdS$_3$ geometries\cite{Lozano:2019jza, Lozano:2019emq}. These backgrounds are thought to be the near horizon limit of D2-D4-D6-D8-NS5 brane set-ups \cite{Lozano:2019zvg}, which give rise to two-dimensional $\mathcal{N} = (0,4)$ QFTs. These theories have two sets of linear quivers, with colour groups $N_n, \tilde{N}_n$ and flavour groups $F_n, \tilde{F}_n$, coupled by matter fields. The colours and flavours of these two quivers can not be independent of each other, but have to be chosen so that the gauge anomalies cancel. The condition for this is
\begin{equation}
\tilde{F}_n = 2N_n - N_{n+1} - N_{n-1}, \hspace{2cm}
F_n = 2\tilde{N}_n - \tilde{N}_{n+1} - \tilde{N}_{n-1}.
\end{equation}
Note this is analogous to the consistency condition in eq.(\ref{eq_CFConsistencyCondition}), and our methods from section \ref{subsec_StartingFromFlavours} therefore apply to this case as well.  

The supergravity solution is of the form
\begin{equation}
(AdS_3 \times S^2 \times CY_2 ) \ltimes \mathbb{R}.
\end{equation}
Here we will just give the expression for the metric. In addition this background contain a dilaton, $B_2$-field and RR $F_0$, $F_2$, and $F_4$-flux.
\begin{eqnarray}
&ds^2 =& \frac{u}{\sqrt{\hat{h}_4h_8}}\left( ds_{AdS_3}^2 + \frac{\hat{h}_4 h_8}{4\hat{h}_4 h_8 + (u')^2}d\Omega^2_2 \right) + \sqrt{\frac{\hat{h}_4}{h_8}}ds^2_{CY_2} + \frac{\sqrt{\hat{h}_4 h_8}}{u}dz^2
\end{eqnarray}
As in the other cases, the background is completely determined in terms of the functions $u(z)$, $\hat{h_4}(z)$ and $h_8(z)$, that have to be at most linear in $z$,
\begin{equation}
u''=0, \hspace{2cm}
h_8''=0, \hspace{2cm}
\hat{h}_4''=0.
\end{equation}
The functions $\hat{h}_4(z)$ and $h_8(z)$ are directly related to the rank functions of these two quivers as
\begin{equation}
\hat{h}_4(z) = R(z), \hspace{1cm}
h_8(z) = \tilde{R}(z).
\end{equation}

\section{Conclusions and Future Directions}\label{sec_Conclusion}
We have shown how to make sense of continuous rank functions, as describing infinitely long quivers. We showed that these continuous rank functions have well-defined supergravity duals with an infinite number of smeared NS5, D$_p$ and D$_{p+2}$-branes. We calculated the Page and central charges corresponding to these backgrounds with continuous rank functions and showed they agree with the results obtained for infinitely long discrete rank functions. 

These continuous rank functions generalise the supergravity solutions that one normally considers to be the dual descriptions of quiver field theories. Since the supergravity description is valid for very large quivers, it might be more convenient to think of all continuous rank functions as valid descriptions of dual SCFTs.

We demonstrated this explicitly in the context of the half-BPS AdS$_5$ and AdS$_7$ geometries, where we constructed examples of a parabolic and sinusoidal quiver. Here these infinite quivers with continuous rank function $R(z) \in C^{(\infty)}$ can be defined by fixing their flavour groups and fixing the colours by requiring that consistency conditions are satisfied. 

Certain CFTs that are dual to supergravity solutions with these continuous rank functions might have particularly nice properties. A recently found example of this is a sinusoidal rank function of the form $R(z) = A \sin (\omega z)$, for which the bosonic sector of the string on the corresponding AdS$_7$ geometry is integrable \cite{Filippas:2019puw}. We hope to report on more interesting SCFTs that are dual to supergravity solutions with continuous rank functions in the near future.\\
\\
Performing field theoretical calculations for a `scaled' short quiver, obtained after the scaling procedure in section \ref{subsec_ScalinginiteQuiver} can be quite difficult. It would be very interesting to study these continuous quivers from a field theoretical perspective. For example, the parabolic quiver in section \ref{subsec_ParabolicQuiver} can be defined by setting the ranks of all flavour groups equal to a constant. Similarly, the sinusoidal quiver in section \ref{subsec_SinusoidalQuiver} can be defined by choosing the ranks of all flavour groups proportional to the colour groups. It would be very interesting to see if certain field theoretical calculations for these particular quivers could become feasible, and whether this would allow a more direct study of the field theories that are expected to be dual to these half-BPS supergravity descriptions. 

It would furthermore be interesting to work out an equivalent description of continuous rank functions for the other classes of half-BPS AdS$_4$ and AdS$_6$ geometries. 
In Appendix \ref{App_AdSGeometries} we briefly comment on an AdS$_4$ and an AdS$_6$ geometray that were obtained from a non-Abelian T-duality and have an infinite linear quiver associated with them \cite{Lozano:2016wrs, Lozano:2018pcp}. The functions that characterise these NATD geometries contain smeared branes. It would be interesting to see if these particular solutions could be used as a starting point to define more general half-BPS AdS$_4$ and AdS$_6$ backgrounds corresponding to infinitely long quivers and continuous rank functions in a similar way as we did here for the half-BPS AdS$_7$ and AdS$_5$ geometries.

\section{Acknowledgements}
The author would like to thank Carlos N\'{u}\~{n}ez for many productive discussions, and Yolanda Lozano, Daniel Thompson, Stefano Cremonesi, and Alessandro Tomasiello for valuable comments on the draft of this paper.

JvG acknowledges the support of an STFC studentship under DTP grant ST/N504464/1.

\appendix

\section{Half-BPS AdS$_{4}$ and AdS$_6$ Geometries}\label{App_AdSGeometries}
For the AdS$_{p+1}$ duals of the D$_p$-D$_{p+2}$-NS5 brane set-ups the situation for $p=3$ and $p=5$ is different from $p =4$ and $p=6$ that we have discussed in this paper so far.  The quiver structure of the dual field theories is for these AdS$_4$ and AdS$_6$ geometries encoded in the poles of the functions $\mathcal{F}$ at the boundary of $\Sigma_2$. This makes it not directly obvious how we could define the analogue of a continuous rank function for those cases. 

In addition, the consistency condition that related the ranks of the flavour groups to the second derivative of colour rank function is no longer valid in these cases. Our approach to construct infinite quivers in section \ref{subsec_StartingFromFlavours} by defining the flavour groups does therefore not directly apply here. 

We briefly introduce these cases here, and make some oobservations on infinite linear rank functions that are interpreted solutions with smeared branes \cite{Lozano:2016wrs, Lozano:2018pcp}. We suggest these cases could be used as a starting point to define continuous rank functions for infinite quivers for the AdS$_4$ and AdS$_6$ cases.

\subsection{AdS$_4$}
D3-D5-NS5 brane set-ups give rise to 3d $\mathcal{N} = 4$ supersymmetric, mirror symmetric
gauge theories known as $T^\rho_{\hat{\rho}}(SU(N))$-theories, introduced in \cite{Gaiotto:2008ak}. 
These theories are characterised by two partitions $\rho$ and $\hat{\rho}$ of an integer $N$, that one can relate to a corresponding linear quiver diagram as,
\begin{eqnarray}
&\rho &= \{ \underbrace{\ell^{(1)}, \ldots \ell^{(1)}}_{N_{D5}^{(1)}}, \ldots , \underbrace{\ell^{(n)}, \ldots \ell^{(n)}}_{N_{D5}^{(n)}}, \ldots, \underbrace{\ell^{(p)}, \ldots \ell^{(p)}}_{N_{D5}^{(p)}} \}, \\
&\hat{\rho} &= \{ \underbrace{\hat{\ell}^{(1)}, \ldots \hat{\ell}^{(1)}}_{N_{NS5}^{(1)}}, \ldots , \underbrace{\hat{\ell}^{(n)}, \ldots \hat{\ell}^{(n)}}_{N_{NS5}^{(n)}}, \ldots, \underbrace{\hat{\ell}^{(p)}, \ldots \hat{\ell}^{(p)}}_{N_{NS5}^{(p)}} \}.
\end{eqnarray}
These theories have an IR conformal fixed point when
\begin{equation}
F_n > 2N_n - N_{n+1} - N_{n-1}.
\end{equation}
This now implies the values of the flavour groups have to be larger than the second derivative of the rank function, and thus our approach to construct infinite quivers in section \ref{subsec_StartingFromFlavours} does not directly apply here!\\ 
\\
At the conformal fixed point the symmetry group is enhanced to $OSp(2,2|4)$, and the system is thought to have a Type IIB dual description in terms of 
\begin{equation}
\left( AdS_4\times S^2 \times S^2 \right) \ltimes \Sigma_2,
\end{equation}
where $\Sigma_2$ has the topology of a disk. When we parametrise  $\Sigma_2$ in terms of a complex coordinate $z$, the geometry is given by \cite{DHoker:2007zhm, DHoker:2007hhe}
\begin{eqnarray}
& ds^2 &= 2 \left( \frac{N_1 N_2}{W^2} \right)^{1/4} \left[ ds_{AdS_4}^2 
+ \frac{h_1^2 W}{N_1} ds^2_{\Omega^2_1} 
+ \frac{h_2^2 W}{N_2} ds^2_{\Omega^2_2}
+ \frac{2W}{h_1 h_1}dz d\bar{z} \right] 
\end{eqnarray}
Here we omitted the expressions for the dilaton, $B_2$-field, and $F_3$ and $F_5$ flux as we will not perform any calculations on this geometry. See also \cite{Assel:2011xz, Aharony:2011yc} for extensive reviews.

The geometry is completely specified in terms of two harmonic functions $h_1(z, \bar{z})$ and $h_2(z, \bar{z})$, and their derivatives, with $W = \partial \bar{\partial}( h_1 h_2 )$, $N_1 = 2h_1 h_2 |\partial h_1|^2 -h_1^2 W$ and $N_2 = 2h_1 h_2 |\partial h_2|^2 -h_2^2 W$. These functions only depends on $\Sigma_2$, that in these coordinates corresponds to a strip of the upper-half complex plane
\begin{equation}
\Sigma = \{ z \in \mathbb{C}\;|\;0 \leq \text{Im}\;z \leq \frac{\pi}{2} \}, \nonumber
\end{equation}
with the two sphere vanishing at $\text{Im}\;z =0, \frac{\pi}{2}$. The solutions for $h_1$ and $h_2$ can be written as a sum over harmonic functions of the form
\begin{eqnarray}
&& h_1 = \left[ -i\alpha_1 \sinh(z-\beta_1) - \gamma_1 \ln \tanh \left( \frac{i\pi}{4} - \frac{z-\delta_1}{2}\right) \right] + c.c. \nonumber \\
&& h_2 = \left[ \alpha_2 \cosh(z-\beta_2) - \gamma_2 \ln \tanh \left(  \frac{z-\delta_2}{2}\right) \right] + c.c. \nonumber
\end{eqnarray}
with the $( \alpha_i, \beta_i, \gamma_i, \delta_i )$ real parameters. The quiver structure of the field theory is now encoded in the poles on the upper and lower boundary of $\Sigma_2$ of these functions, with the residues of the poles of $h_1$ and $h_2$ giving respectively the NS5 and D5 fluxes.\\ 
\\
The quiver structure of the field theory is now encoded in the poles of these harmonic functions $h_1$ and $h_2$ at the boundary of $\Sigma_2$. This makes it less obvious how we could obtain the geometry for a continuous quiver with infinite NS5 and D5-brane charges, where we would expect these poles to somehow become smeared.

Interestingly, in \cite{Lozano:2016wrs} it was shown that the non-Abelian T-dual of a particular reduction of AdS$_4 \times$S$^7$ to ten dimensions fits within this class of $\frac12$BPS AdS$_4$ solutions. The corresponding geometry has an infinite linear quiver with an infinite number of NS5-branes, and linearly increasing numbers of D3 and D5-branes (similar to how the NATD of AdS$_5 \times$S$^5$ \cite{Sfetsos:2010uq} fits in the $\frac12$ BPS class of AdS$_5$ geometries and also has an infinite linear quiver \cite{Lozano:2016kum}). As a consequence the functions $h_1$ and $h_2$ obtained for this geometry do not have poles, but instead have a continuum of smeared NS5 and D5-branes at their boundary. The resulting functions are of the form
\begin{equation}
 h_1 = \frac{kr(1+\sigma)}{4\beta}, \hspace{2cm}
 h_2 = \frac{1-\sigma}{2\beta}, \nonumber
\end{equation}
with $\sigma$ and $r$ respectively related to the real and imaginary parts of $z$. It would be very interesting to see if this particular solution could be used as a starting point to define backgrounds with infinitely long quivers and continuous rank functions for these AdS$_4$ geometries.

\subsection{AdS$_6$}
The discussion we will present here for the AdS$_6$ geometries follows  the lines of the previous section. The D5-D7-NS5 brane set-ups are a bit more complicated. Charge conservation does not allow D5-brane's to end on an NS5-brane, and at the intersection point, the branes will instead attract each other and merge to form a (p,q) 5-brane, which allows the formation of so-called (p,q)-5-brane webs \cite{Aharony:1997ju, Aharony:1997bh}. 

We can still consider a particular brane-web that arises from a linear quiver diagram, but the condition for these theories to have a conformal fixed point will now be \cite{Intriligator:1997pq, Bergman:2015dpa}
\begin{equation}
F_n \leq 2N_n + 4.
\end{equation}
This implies the values of the flavour groups are no langer related to the second derivative of the rank function, and our approach to construct infinite quivers in section \ref{subsec_StartingFromFlavours} does not apply anymore!\\
\\
These set-ups give rise to 5d $\mathcal{N} = 1$ supersymmetric QFTs. At the conformal fixed point the symmetry group is enhanced to $F(4)$ (the unique superconformal group in 5 dimensions). It is thought these SCFTs have a dual description in terms of a Type IIB geometry of the form
\begin{equation}
\left( AdS_6 \times S^2 \right) \ltimes \Sigma_2
\end{equation}
Where $\Sigma_2$ has the topology of a disk. We can again choose coordinates where $\Sigma_2$ is parameterised by a complex coordinate $z$. These Type IIB backgrounds with an $F(4)$ isometry group were given in \cite{DHoker:2016ujz, DHoker:2017mds, DHoker:2017zwj}. 
\begin{eqnarray}
& ds^2 &= f_6^2\;ds_{AdS_6}^2 + f_2^2\;d\Omega_2^2 + ds_\Sigma^2
\end{eqnarray}
Here we largely omit the details of these backgrounds as their expressions are rather complex. See \cite{DHoker:2016ysh} for a shorter review. These AdS$_6$ backgrounds have an NS-NS sector with dilaton and $B_2$-flux, the RR-RR sector consists of an $F_3$ and $F_5$-flux. 

As before, all functions of the geometry are completely determined in terms of two locally holomorphic functions $\mathcal{A}_\pm$ that only depend on the complex coordinate $z$. 
Analogous to the earlier discussed AdS$_4$ backgrounds, the quiver structure will be encoded in terms of the poles of these functions $\mathcal{A}_\pm$ on $\Sigma_2$. The charges of the $(p,q)$ 5-branes will correspond to the residues of these poles, while the D7-branes are related to punctures of $\Sigma_2$. 
As in the AdS$_4$ case this makes it less obvious how we could obtain the geometry for a continuous quiver with an infinite number of NS5 and D-branes.\\
\\
Interestingly, it was shown in \cite{Lozano:2018pcp} that the non-Abelian T-dual of the AdS$_6 \ltimes$S$^4$ Brandhuber-Oz background lies in this class of $\frac12$BPS AdS$_6$ solutions and has again a linear quiver associated with it. The corresponding holomorphic functions $\mathcal{A}_\pm$ for this solution have smeared NS5 and D7/O7-branes on the boundary of $\Sigma$.
It would be very interesting to see if this particular solution could be used as a starting point to define backgrounds with infinitely long quivers and continuous rank functions for these AdS$_6$ geometries.

\section{Expressions for General Quivers in 4d and 6d}\label{sec_GeneralExpressions}
Let us consider the most general quiver field theory. The gauge group is 
\begin{equation}
SU(N_1)\times SU(N_2)\times....\times SU(N_P).\nonumber
\end{equation}
The flavour group, with each element associated with a colour node, is
\begin{equation}
SU(F_1)\times SU(F_2)\times....\times SU(F_P).\nonumber
\end{equation}
The consistency condition for the 4d and 6d SCFTs is,
\begin{equation}
2N_k- N_{k+1}-N_{k-1}= F_k.\label{condxxx}
\end{equation}
We are using that $N_{-1}=N_{P+1}=0$. The rank function is,
  \begin{equation} \label{profilexx}
R(z)
                    =\left\{ \begin{array}{ccrcl}
                       N_1 z & 0\leq z\leq 1 \\
                       N_1+(N_2-N_1)(z-1) &1 \leq z \leq 2 \\
                        N_2 + (N_3-N_2)(z-2) & 2\leq z \leq 3\\
                        N_k +(N_{k+1}-N_k)(z-k) &~~ k\leq z \leq k+1,\;\;\;\; k:=3,....,P\\
                        N_P - N_P(z-P) & P\leq z\leq P+1.
                                             \end{array}
\right.
\end{equation}

\subsection{General Expressions for AdS$_5$ Gaiotto-Maldacena}
In the Gaiotto-Maldacena AdS$_5$ geometry of eq.(\ref{eq:10dGaiottoMaldacenaNS}), the rank function $R(z) = \lambda(z)$. We can then calculate the number of branes in the corresponding Hanany-Witten set-up using the expressions proposed in \cite{Nunez:2019gbg}.
\begin{eqnarray}
& & N_{D4}= \int_{0}^{P+1} \lambda(z) dz = 
N_1+N_2+....+N_P,\;\;\;\;\; N_{NS5}=P+1,\nonumber\\
& & N_{D6}= \left( \lambda'(0)-\lambda'(P+1)\right)= N_1+N_P=F_1+F_2+....+F_P.\label{GMcargasgood}
\end{eqnarray}
In addition, we can derive a general expression for the central charge of the quiver, using the expression derived in \cite{Nunez:2019gbg}.
\begin{equation}\label{eq_MGCentralCharge}
c = \frac{2}{\pi^4} \int_0^{P+1}\lambda^2(z)\;dz
\end{equation}
Evaluating this expression for the general rank function gives 
\begin{equation}
\frac{\pi^4}{2} c = \sum_{j=1}^{P} \left( N_j \sum_{i=1}^{P} N_i \right)
\end{equation}

\subsection{General Expressions for AdS$_7$ Cremonesi-Tomasiello}
In the Cremonesi-Tomasiello AdS$_7$ geometry of eq.(\ref{eq:TomasielloGeometryGeneral}), the rank function $R(z)=-\frac{\alpha''(z)}{81\pi^2}$. We can write from here the function $\alpha(z)$. Imposing that $\alpha(z=0)=0$, we find after some algebra
\begin{equation} \label{profilealpha(z)}
-\frac{\alpha(z)}{81\pi^2}
                    =\left\{ \begin{array}{ccrcl}
                      b_0 z+ N_1\frac{ z^3}{6} & 0\leq z\leq 1 \\
                  a_k+ b_k(z-k) + \frac{ N_k}{2}(z-k)^2+\frac{(N_{k+1}-N_k)}{6}(z-k)^3 &\;\;\;\;k \leq z \leq (k+1),\;\;\;\; k:1,....,P-1 \\
   a_p + b_p(z-P)+\frac{ N_P}{2}(z-P)^2 - \frac{N_P}{6}(z-P)^3 & P\leq z\leq P+1.
                                             \end{array}
\right.
\end{equation}
To satisfy the conditions of  continuity for $\alpha(z)$ and of $\alpha'(z)$ and also imposing $\alpha(z=P+1)=0$ leads to the coefficients,
\begin{eqnarray}
& & b_0=-\frac{1}{(P+1)}\sum_{i=1}^P (P+1-i)N_i,\label{coefficients}\\
& & a_k=k b_0 +\frac{N_k}{6}+\sum_{i=1}^{k} (k-i)N_i ,\;\;\;\;\; k:1,....P,\nonumber\\
& & b_k= b_0+\frac{N_k}{2}+\sum_{i=1}^{k-1} N_i, \;\;\;\;\;\; k:=1,....,P.\nonumber
\end{eqnarray}
From here we find the number of branes in the associated Hanany-Witten set-up. Using the expressions proposed in \cite{Filippas:2019puw}, we find 
\begin{eqnarray}
& & N_{D6}=-\frac{1}{81\pi^2}\int_{0}^{P+1} \alpha''(z) dz=N_1+N_2+....+N_P,\;\;\;\;\; N_{NS5}=P+1,\nonumber\\
& & N_{D8}=\frac{1}{81\pi^2}\left( \alpha'''(0)-\alpha'''(P+1)\right)=N_1+N_P=F_1+F_2+....+F_P.\label{cargasgood}
\end{eqnarray}
More interestingly, we can derive a general expression for the central charge of the quiver, using the expression derived in \cite{Nunez:2018ags}.
In fact, in general we have,
\begin{equation}\label{eq_TCCentralCharge}
c=-\frac{2^8}{3^8\times 16\times G_N}\int_{0}^{P+1} \alpha''(z)\alpha(z) dz,\;\;\;\;\; G_N=8\pi^6.
\end{equation}
Evaluating this expression in general gives us
\begin{eqnarray}
& & -\frac{\pi^2}{2} c= \left[\frac{b_0 N_1}{3} +\frac{N_1^2}{30}  \right] +\left[\frac{a_P N_P}{2} +\frac{b_P N_P}{6} +\frac{N_P^2}{30}   \right] +\nonumber\\
& &+ \frac{1}{30}\sum_{k=1}^{P-1}\left[N_k^2 +3 N_k N_{k+1} + N_{k+1}^2 + 15 a_k(N_k +N_{k+1}) + 5 b_k(N_k+2 N_{k+1})      \right].\label{genericcc}
\end{eqnarray}

\subsection{Two Examples}
We can test the expression in eq.(\ref{genericcc}) in two particularly simple cases:

First, in the case in which all the ranks are equal $N_i=N$. This needs of only two flavour groups, of rank $N$ at the beginning and at the end of the quiver. In this case, calculating the expressions in eqs.(\ref{coefficients}),(\ref{genericcc}), we find
\begin{eqnarray}
& & b_0=-\frac{NP}{2},\;\;\; a_k=\frac{N}{2}(-kP +k^2-k+\frac{1}{3}),\;\;\; b_k=\frac{N}{2}(2k-P-1), \;\; k:1,....,P.\nonumber\\
& & c=\frac{N^2 P^3}{6\pi^2}(1+\frac{3}{P} -\frac{1}{P^2} +\frac{1}{5P^3})\sim \frac{N^2 P^3}{6\pi^2}.\label{central1}
\end{eqnarray}
This last result makes sense, because the supergravity approximation is good when $P\to\infty$, that is in the limit of long-linear quivers. In this case, the central charge calculated above  precisely with that derived in \cite{Nunez:2018ags} (see eq.(2.16) of that paper).
\begin{figure}[t]
{
 \centering
 \subfloat[\small ]{
     \includegraphics[width=0.45\textwidth]{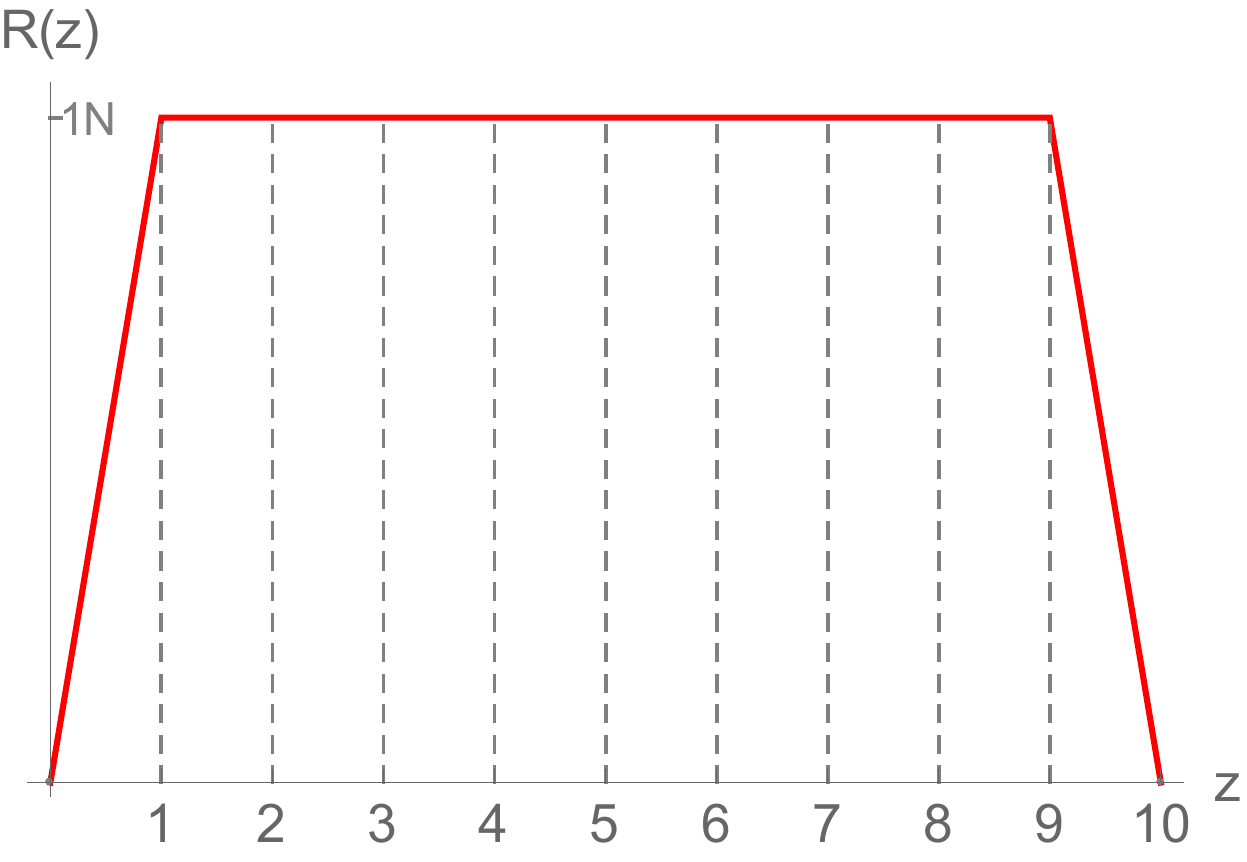}}\hspace{0.05\textwidth}
 \subfloat[\small ]{
    \includegraphics[width=0.45\textwidth]{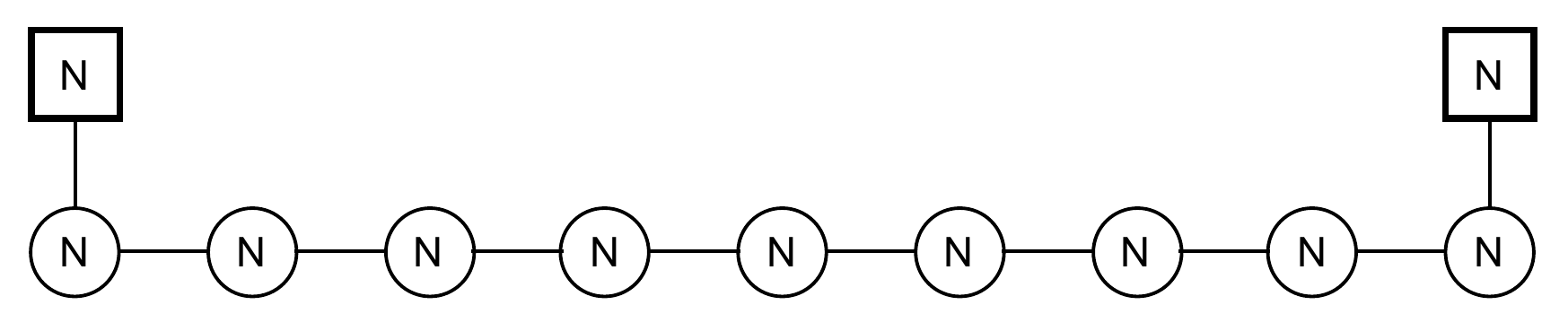}}   
\caption{An example of the rank function (a) and quiver (b) for the first case with all the ranks equal $N_i = N$, and two flavour groups, of rank $N$ at the beginning and end of the quiver.}
}
\end{figure}\\
\\
A second test is to consider the quiver in which the ranks increase as $N_k=k N$. This only needs of a flavour group, or rank $(P+1)$ attached to the last node of rank $PN$.
The expressions in eqs.(\ref{coefficients}),(\ref{genericcc}), read in this case
\begin{eqnarray}
& & b_0=-\frac{NP}{6}(P+2),\;\;\; a_k=\frac{kN}{6}(-2P +k^2-P^2),\;\;\; b_k=\frac{N}{6}(3 k^2-P^2-2P), \;\; k:1,....,P.\nonumber\\
& & c=\frac{2N^2 P^5}{45\pi^2}(1+\frac{5}{P} +\frac{5}{P^2} +\frac{1}{P^3})\sim \frac{2N^2 P^5}{45\pi^2}.\label{central2}
\end{eqnarray}
In the limit of long-linear quivers the central charge calculated above  precisely with that derived in \cite{Nunez:2018ags}, see eq.(2.15) in that paper. We have then good reasons to believe that the expression in eq.(\ref{genericcc}) is correct.
\begin{figure}[t]
{
 \centering
 \subfloat[\small ]{
     \includegraphics[width=0.45\textwidth]{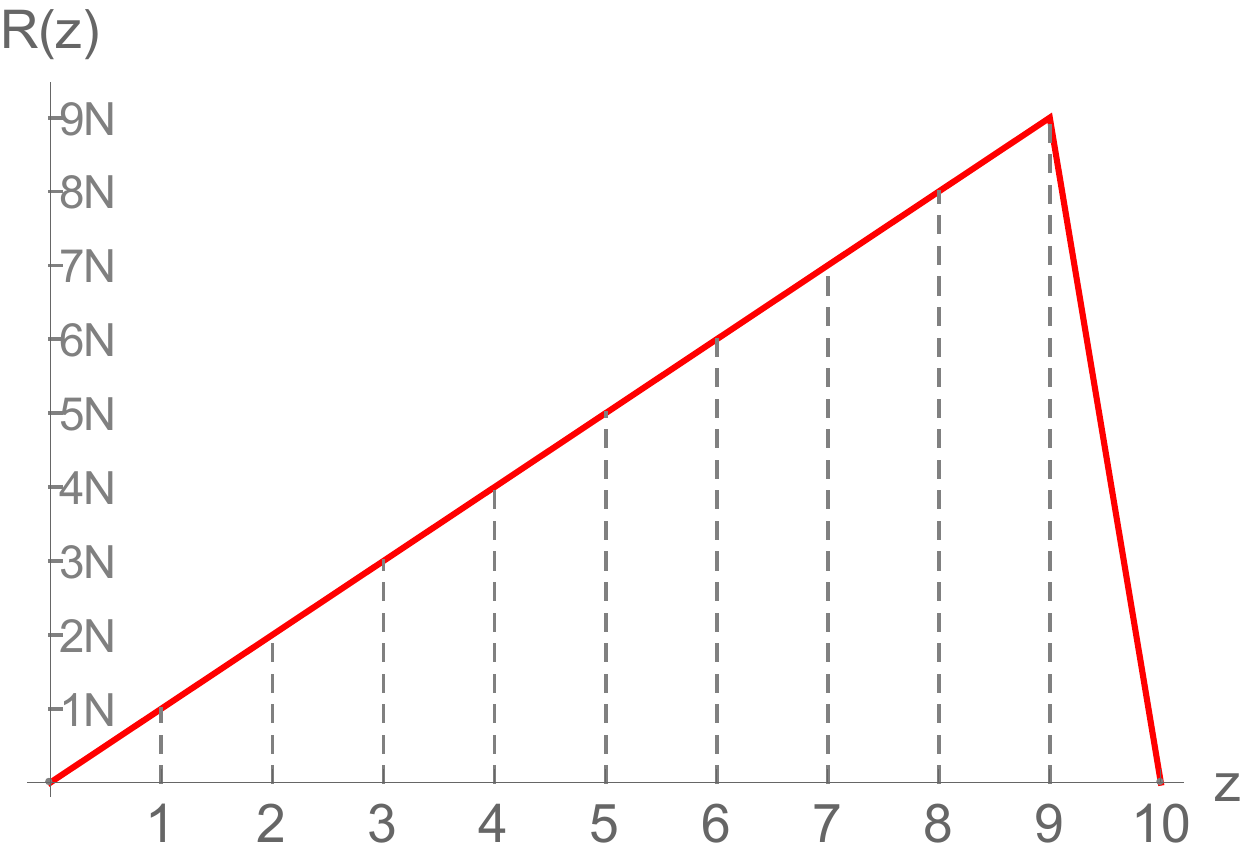}}\hspace{0.05\textwidth}
 \subfloat[\small ]{
    \includegraphics[width=0.45\textwidth]{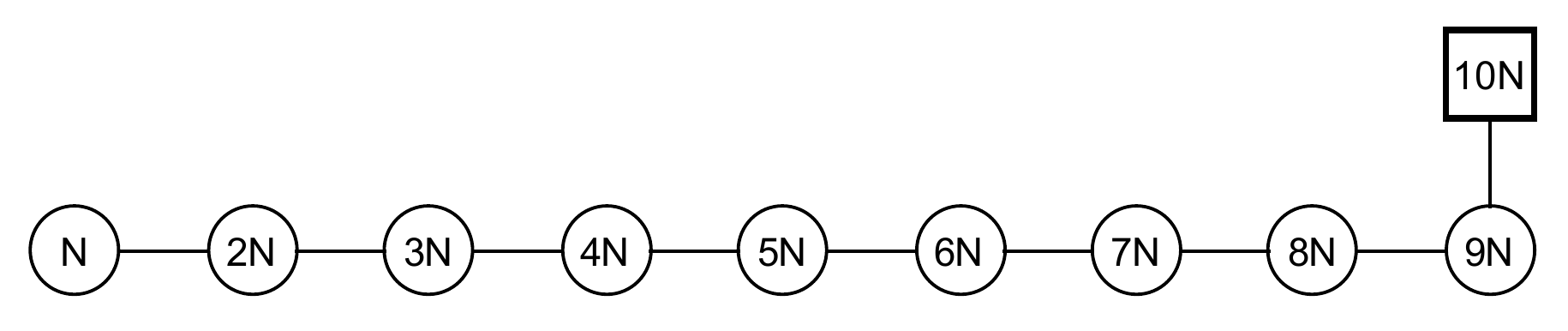}}   
\caption{An example of the rank function (a) and quiver (b) for the second case with all the ranks linearly increasing as $N_k = k N$, and a flavour group of rank $(P+1)$ at the end of the quiver.}
}
\end{figure}\\
\\

 \end{document}